\documentclass[12pt]{iopart}
\usepackage{xcolor}
\usepackage{graphicx}
\usepackage{dcolumn}
\usepackage{bm}
\usepackage{orcidlink}
\usepackage{multirow}
\usepackage[numbers,sort&compress,square]{natbib}
\usepackage[numbers,sort&compress,square]{natbib}
\usepackage{hyperref}
\usepackage{subcaption}
\usepackage{verbatim}
\usepackage{soul}
\usepackage{iopams} 
\usepackage{acronym}
\usepackage[mathlines,running]{lineno}
\usepackage{tabularx}
\usepackage{listings}

\hypersetup{
    colorlinks=true,
    linkcolor=blue,
    filecolor=magenta,      
    urlcolor=cyan,
}

\usepackage{tikz}
\usetikzlibrary{shapes.geometric, arrows}

\usepackage[mathlines,running]{lineno}
\begin{document}
\newcommand{\REF}[1]{\textcolor{red}{REFERENCE #1}}

\newcommand{\dmtdqvector}{{\sc DMT-DQ vector}\xspace}
\newcommand{\calibstate}{{\sc GDS-CALIB\_STATE vector}\xspace}
\newcommand{\dqr}{{\sc Data Quality Report}\xspace}
\newcommand{\idq}{i{\sc DQ}\xspace}
\newcommand{\ovl}{{\sc OVL}\xspace}
\newcommand{\qscan}{{\sc Q-scan}\xspace}
\newcommand{\hveto}{{\sc HVeto}\xspace}
\newcommand{\pointy}{{\sc pointy}\xspace}
\newcommand{\ldvw}{{\sc LIGO DataViewer Web}\xspace}
\newcommand{\lvalert}{{\sc LIGO-Virgo Alert System}\xspace}
\newcommand{\gracedb}{{\sc Gravitational-Wave Candidate Event Database}\xspace}
\newcommand{\gwpy}{{\sc GWpy}\xspace}
\newcommand{\gwdetchar}{{\sc GW-DetChar}\xspace}
\newcommand{\gwsumm}{{\sc GWSumm}\xspace}

\newcommand{\dd}[1]{\textcolor{violet}{#1}}

\newcommand{\virgo}{{Virgo}\xspace}

\newcommand{\pycbc}{{\sc PyCBC}\xspace}
\newcommand{\gstlal}{{\sc GstLAL}\xspace}
\newcommand{\cwb}{{\sc cWB}\xspace}

\newcommand{\gwcelery}{{\sc GWCelery}\xspace}

\acrodef{CBs}{cryo-manifold baffles}
\acrodef{CB}{cryo-manifold baffle}
\acrodef{CBC}{Compact binary coalescence}
\acrodef{DetChar}{Detector Characterization}
\acrodef{PEM}{Physical Environment and
Monitoring}
\acrodef{IFO}{interferometer}
\acrodef{LHO}{LIGO Hanford}
\acrodef{LLO}{LIGO Livingston}
\acrodef{GW}{gravitational-wave}
\acrodef{LIGO}{Laser Interferometer Gravitational-Wave Observatory}
\acrodef{BBH}{binary black hole}
\acrodef{O1}{first observing run}
\acrodef{O2}{second observing run}
\acrodef{DQR}{Data Quality Report}
\acrodef{O3}{third observing run}
\acrodef{O3a}{first half of the third observing run}
\acrodef{O3b}{second half of the third observing run}
\acrodef{O4}{fourth observing run}
\acrodef{BH}{black hole}
\acrodef{BBH}{binary black hole}
\acrodef{IMBH}{intermediate-mass black hole}
\acrodef{IMC}{input mode cleaner}
\acrodef{HEPI}{hydraulic external pre-isolator}
\acrodef{SNR}{signal-to-noise ratio}
\acrodef{BNS}{binary neutron star}
\acrodef{PSD}{power spectral density}
\acrodef{GR}{general relativity}
\acrodef{FAR}{false-alarm rate}
\acrodef{GCN}{the Gamma-ray Coordinates Network}
\acrodef{CBC}{compact binary coalescence}
\acrodef{VT}{volume-time}
\acrodef{ASD}{amplitude spectral density}
\acrodefplural{ASD}{amplitude spectral densities}
\acrodef{DAC}{digital-to-analog}
\acrodef{GWB}{gravitational-wave background}
\acrodef{DQ}{data quality}
\acrodef{RRT}{rapid response team}
\acrodef{GRB}{gamma-ray burst}
\acrodef{DARM}{differential arm readout measurement}
\acrodef{OSEM}{optical shadow sensors and magnetic actuator}
\acrodef{oplev}{optical lever}
\acrodef{ETM}{end test mass}
\acrodef{ETMY}{end test mass at the Y-end}
\acrodef{ETMX}{end test mass at the X-end}
\acrodef{AERM}{annular end reaction mass}
\acrodef{LSC}{length sensing and control}
\acrodef{PSL}{pre-stabilized laser}
\acrodef{FSS}{frequency stabilization system}
\acrodef{HAM3}{third horizontal access module}
\acrodef{L2}{penultimate}
\acrodef{L3}{third}
\acrodef{ESD}{electrostatic drive}
\acrodefplural{ESD}{electrostatic drives}
\acrodef{RC}{reaction chain}
\acrodef{ETG}{event trigger generator}
\acrodef{SQZ}{squeezer}
\acrodef{RMS}{root-mean-square}
\acrodef{RF}{radio frequency}
\acrodef{LED}{light emitting diode}
\acrodef{ASC}{alignment sensing and control}
\acrodef{TMS}{transmission motor stage}
\acrodef{OMC}{output mode cleaner}
\acrodef{DAQ}{data acquisition}
\acrodef{FFT}{fast Fourier transform}
\acrodef{ETG}{event trigger generator}
\acrodef{PRC}{Power Recycling Cavity}
\acrodef{SRC}{Signal Recycling Cavity}
\acrodef{AERM}{annular end reaction mass}
\acrodef{QPD}{quadrant photo-diode}
\acrodef{ACB}{Arm cavity baffle}
\acrodefplural{ACB}{arm cavity baffles}

\acrodef{CW}{continuous gravitational-wave}
\acrodef{gws}{gravitational-wave strain}
\acrodef{IGWN}{International Gravitational-wave Network}
\acrodef{LSC}{LIGO Scientific Collaboration}
\acrodef{LVK}{LSC-Virgo-KAGRA Collaboration}
\acrodef{X}{x}
\acrodef{$h(t)$}{gravitational-wave strain timeseries}
\acrodef{GWOSC}{Gravitational-wave Open Science Center}
\acrodef{HVAC}{heating, ventilation, and air conditioning}
\title{LIGO Detector Characterization in the first half of the fourth Observing run.}
\author{%
S.~Soni\,
\orcidlink{0000-0003-3856-8534}$^{1}$,
B.~K.~Berger\,\orcidlink{0000-0002-4845-8737}$^{2}$,
D.~Davis$^{3}$,
F.~Di.~Renzo\,\orcidlink{0000-0002-5447-3810}$^{4}$,
A.~Effler\,\orcidlink{0000-0001-8242-3944}$^{5}$, 
T.~A.~Ferreira$^{6}$,
J.~Glanzer\,\orcidlink{0009-0000-0808-0795}$^{6}$,
E.~Goetz\,\orcidlink{0000-0003-2666-721X}$^{7}$,
G.~González$^{6}$,
A.~Helmling-Cornell\,\orcidlink{0000-0002-7709-8638}$^{8}$
% ...
}%
\medskip
\address {$^{1}$LIGO, Massachusetts Institute of Technology, Cambridge, MA 02139, USA }
\address {$^{2}$Stanford University, Stanford, CA 94305, USA }
\address {$^{3}$LIGO, California Institute of Technology, Pasadena, CA 91125, USA }
\address{$^{4}$Université Lyon, Université Claude Bernard Lyon 1, CNRS, IP2I Lyon IN2P3, UMR 5822, F-69622 Villeurbanne, France}
\address {$^{5}$LIGO Livingston Observatory, Livingston, LA 70754, USA }
\address {$^{6}$Louisiana State University, Baton Rouge, LA 70803, USA }
\address {$^{7}$University of British Columbia, Vancouver, BC V6T 1Z4, Canada }
\address {$^{8}$University of Oregon, Eugene, OR 97403, USA }
% ...
\ead{siddharthsoni22@gmail.com}

\begin{abstract}Progress in gravitational-wave astronomy depends upon having sensitive detectors with good data quality. Since the end of the LIGO-Virgo-KAGRA third Observing run in March 2020, detector-characterization efforts have lead to increased sensitivity of the detectors, swifter validation of gravitational-wave candidates and improved tools used for data-quality products. In this article, we discuss these efforts in detail and their impact on our ability to detect and study gravitational-waves. These include the multiple instrumental investigations  that led to reduction in transient noise, along with the work to improve software tools used to examine the detectors data-quality. We end with a brief discussion on the role and requirements of detector characterization as the sensitivity of our detectors further improves in the future Observing runs. 

\end{abstract}


\section{Introduction}\label{section_intro}

The \ac{LIGO}~\cite{LIGOScientific:2014pky} and Virgo~\cite{VIRGO:2014yos} detectors started the era of \ac{GW} astronomy when the first \ac{GW} signal from the merger of two black holes was detected in 2015~\cite{PhysRevLett.116.061102}. Since then, the \ac{LIGO}, Virgo and KAGRA collaborations have published 90 probable detections of signals involving black holes and neutron stars, including the spectacular multi-messenger discovery in August 2017~\cite{LIGOScientific:2018mvr, LIGOScientific:2021usb, KAGRA:2021vkt, LIGOScientific:2017vwq}. The data used for these analyses is publicly available in the Gravitational Wave Open Science Center (GWOSC)
\footnote{\href{https://gwosc.org}{www.gwosc.org}}. 
The \ac{LIGO} and KAGRA detectors started taking data in the \ac{O4} on May 24, 2023. The first part of the run, O4a, finished on January 16, 2024. The \ac{LIGO} and Virgo detectors started taking data again on April 10, 2024 (O4b). In O4a, 92 significant detection candidates were shared as public alerts, including 11 retracted for data-quality problems. The 81 significant detections are almost as many as the 90 detections published in the GWTC-3 catalog~\cite{KAGRA:2021vkt} from the previous three Observing runs. 

\begin{figure}[t]
    \centering
    \includegraphics[width=\textwidth]{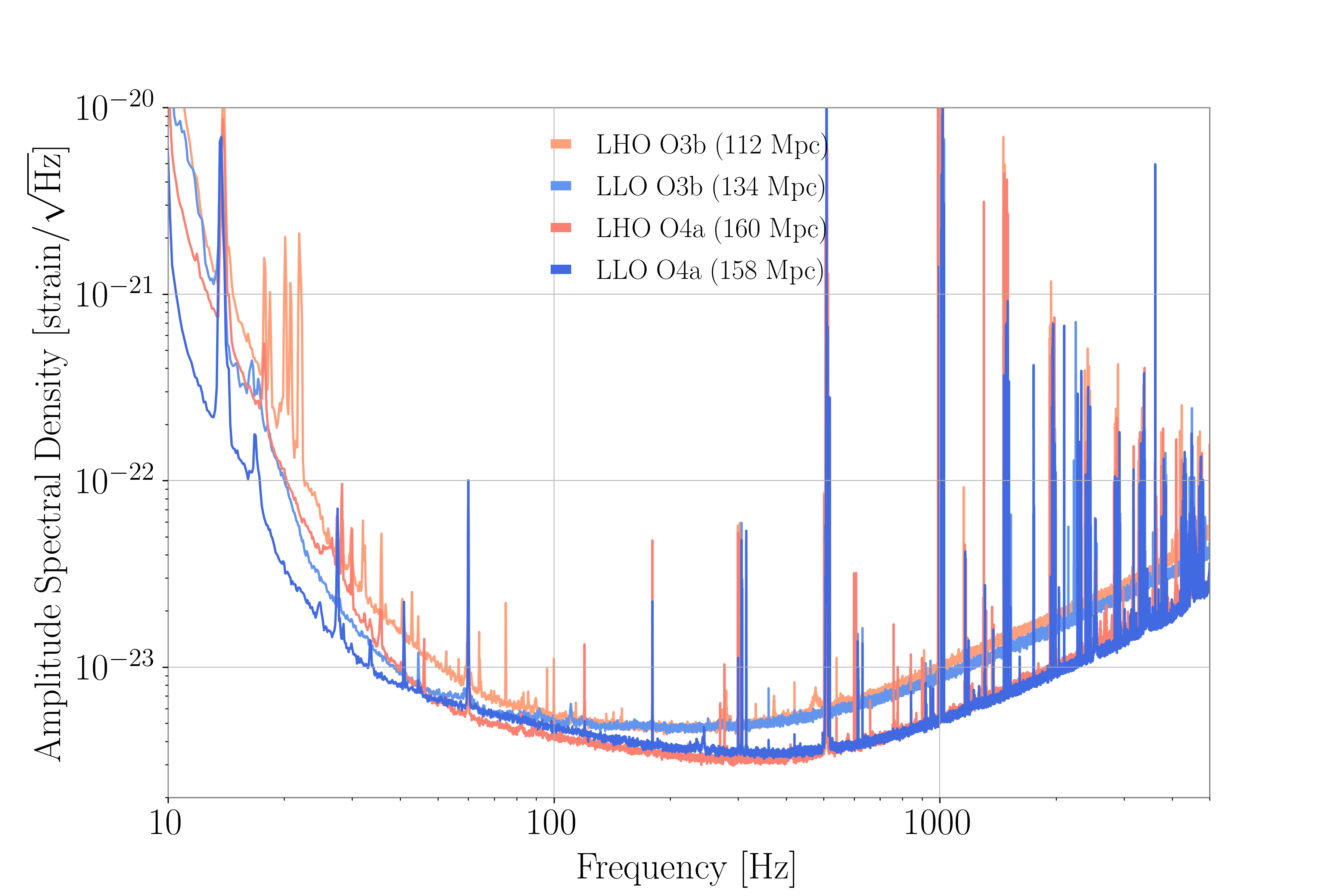}
    \caption{Typical amplitude spectral densities of strain noise at \ac{LHO} and \ac{LLO} detectors in O3b and O4a. The notable improvement in broadband noise reduction observed in O4a is due to various factors including the reduction of scattered-light noise, commissioning of feedback control loops, and the implementation of frequency-dependent squeezing. At \ac{LLO}, the strain noise near 100~Hz is slightly higher than at \ac{LHO}, possibly due to higher thermal noise in end test masses replaced at \ac{LLO}. At low frequencies, the strain noise in both detectors was significantly improved with the removal of a septum window that was coupling ground vibration through scattered light. The damping of baffle resonances also reduced the coupling of scattered light. At \ac{LHO}, an alignment dither system that was used in O3 and produced narrow features around 20~Hz was replaced by a camera servo to avoid the appearance of those lines.}
    \label{O3bO4aASDs_fig}
\end{figure}

As shown in \fref{O3bO4aASDs_fig}, the broadband sensitivity of the \ac{LIGO} detectors in O4a was significantly better than in O3b. Some key upgrades commissioned between \ac{O3} and O4 include frequency-dependent squeezing~\cite{LIGOO4Detector:2023wmz}, replacement of some optics that had small defects (point absorbers) allowing for increased laser power, and replacing squeezing subsystem Faraday isolators with more efficient ones to increase the level of squeezing achieved~\cite{LIGOScientific:2024elc}. 
As a measure of sensitivity, we can calculate the distance to which the noise would allow a detection of a \ac{BNS} system (1.4 M$_\odot$ each) with a signal-to-noise ratio of 8; this is often called the ``binary neutron star inspiral range" or ``BNS range". The spectra shown in \fref{O3bO4aASDs_fig} correspond to dates in O3b (\ac{LHO} March 19, 2020, and \ac{LLO} January 4, 2020) with \ac{BNS} ranges 112 Mpc and 134 Mpc respectively; and dates in O4a (\ac{LHO} December 12 2023, and \ac{LLO}December 31 2023) with \ac{BNS} ranges 160 Mpc and 158 Mpc respectively.

At the junction between instrument scientists and data analysis groups, the \ac{DetChar} group works to understand instrumental noise and improve data-quality~\cite{o2o3_detchar}. The group carries out instrumental investigations in collaboration with instrument scientists at the observatory sites, striving to understand and mitigate sources of noise that are not inherent to the detector design and thereby to improve detector performance. The \ac{DetChar} group also provides data-quality information for \ac{GW} searches, which is crucial to avoid or remove noise artifacts, and to evaluate and validate gravitational-wave signals.

 The \ac{DetChar} group carries out a variety of regular activities to monitor noise and provide data-quality information to searches. The \ac{LIGO} \ac{DetChar} summary pages~\cite{duncan_macleod_2024_10999143} provide detector-site and off-site scientists, technicians, and staff an overview of the performance of the \ac{LIGO} detectors and environmental monitors through a series of regularly updated webpages. Apart from the primary gravitational wave channel that records the strain, there are thousands of auxiliary channels monitoring the state of environment and various detector components. The summary pages centralize plots, figures of merit, and links to additional resources for analyses of the \ac{GW} strain data and various detector subsystems and auxiliary channels. During Observing runs, members of the \ac{DetChar} group participate in data-quality shifts, which utilize the summary pages to closely monitor the performance of the detector. Members also engage in event validation shifts in order to assess any data-quality issues near and during the time of \ac{GW} signals~\cite{LIGO:2021ppb}. Finally, the \ac{DetChar} group works together with \ac{GW} analysis efforts to construct data-quality products in order to avoid noise contamination in \ac{GW} analyses (see \sref{data_qual}.)

Data recorded by the \ac{LIGO} detectors can be characterized as generally Gaussian and stationary with non-Gaussian noise appearing in several forms: 1) short-duration artifacts, often referred to as ``glitches'', which are typically broadband in nature; 2) persistent or slowly time-varying, broadband artifacts; 3) persistent or slowly time-varying, narrowband artifacts, often referred to as ``lines''~\cite{LIGOScientific:2019hgc}. Typically, glitches impact transient \ac{GW} searches and parameter estimation of candidate transient \ac{GW} events but do not strongly impact searches for persistent \ac{GW}. Similarly, lines impact persistent \ac{GW} searches but do not strongly impact searches and parameter estimation for transient \ac{GW}. Persistent broadband artifacts impact all types of searches because the artifacts elevate the noise background.

\begin{figure}[t]
    \centering
    \includegraphics[width=0.9\textwidth]{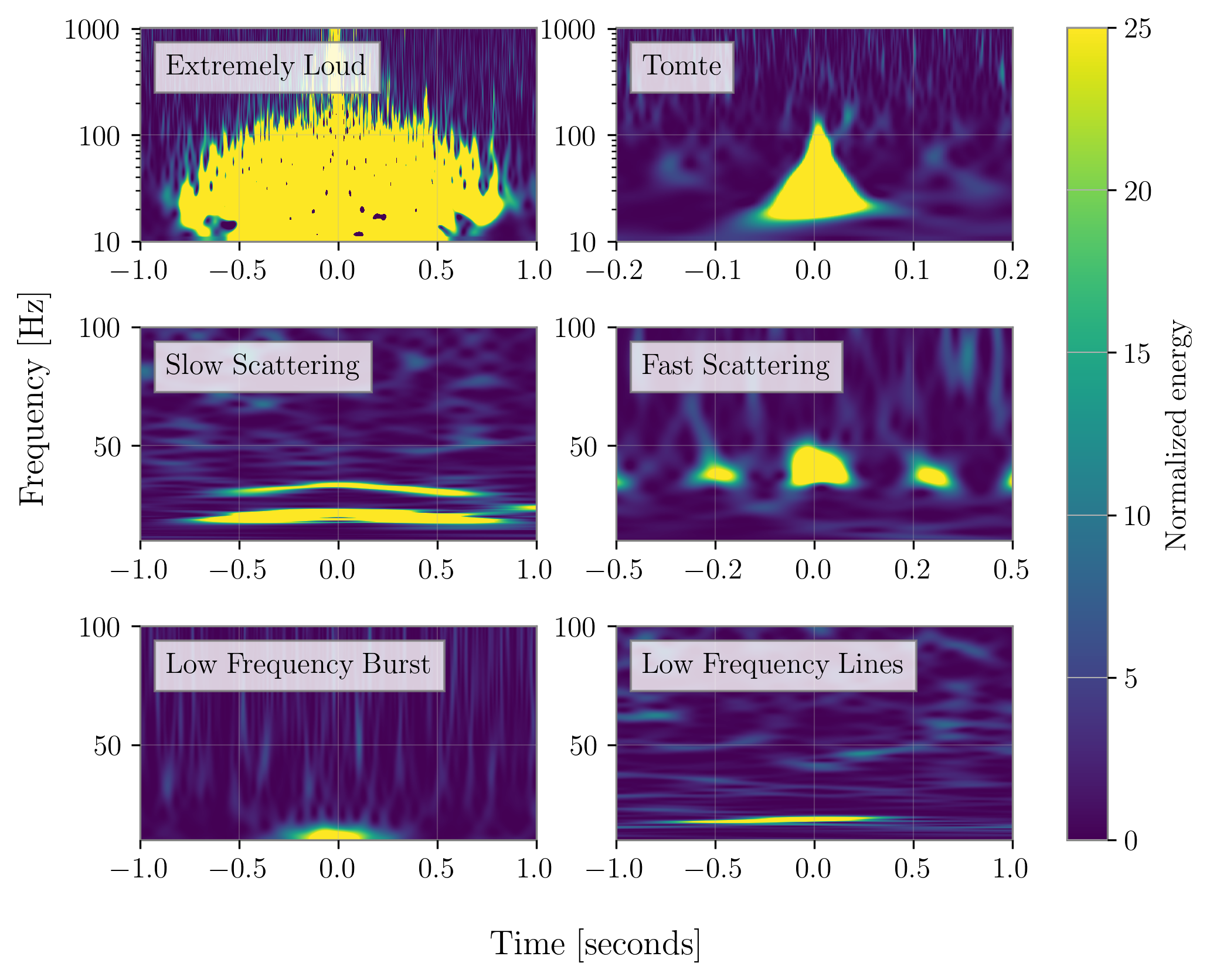}
    \caption{Time-frequency spectrograms of glitches that are common at either LHO or LLO. For Extremely Loud and Tomte, we have not yet found any source. Slow Scattering and Fast Scattering show a strong coupling with increased ground motion and are caused by light scattering \cite{LIGO:2020zwl, Soni:2023kqq}. O4a has seen an increased rate of Low Frequency Burst and Low Frequency Lines at both detectors for which we have not identified a source yet. }\label{common-glitches}
\end{figure}

\begin{figure}[tbh]
    \centering
    \includegraphics[width=\textwidth]{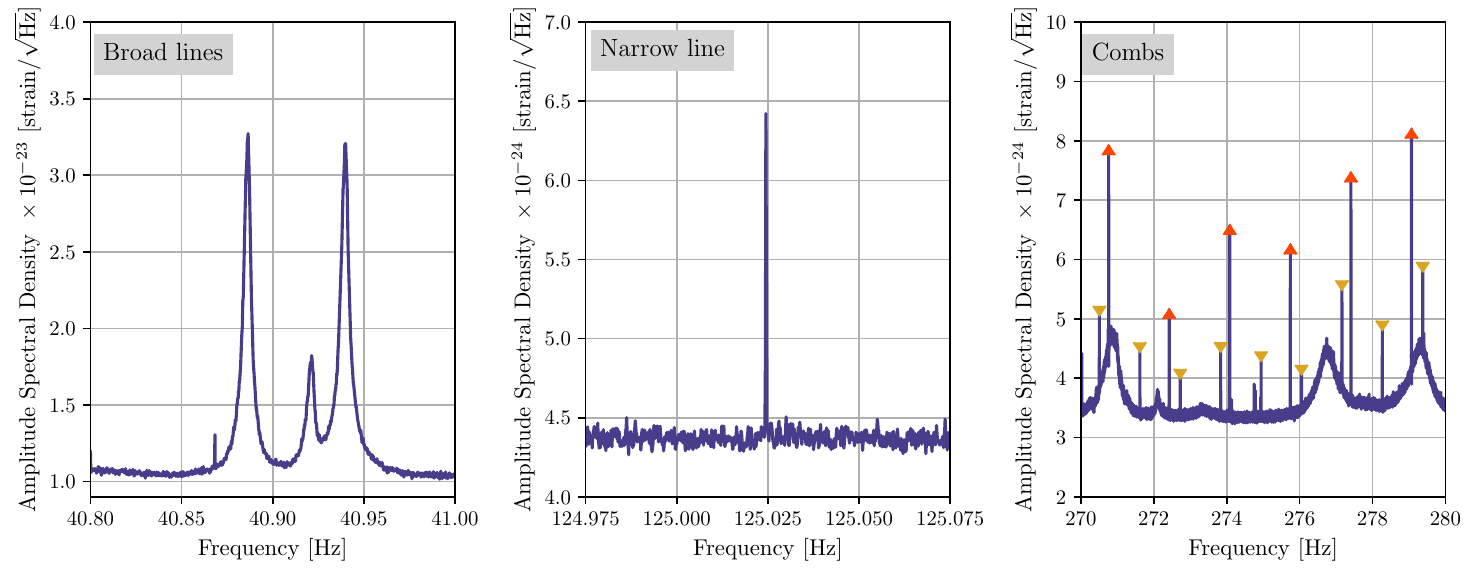}
    \caption{Amplitude spectral density plots of \ac{LHO} strain data showing common line artifact classes that occur at both detectors in high-resolution spectra ($\Delta f = 1/1800$~Hz). Left panel: an example of persistent broad noise artifact arising from interferometer mirror suspension roll-mode resonances. Middle panel: an example of a narrow line of unknown origin. Right panel: an example of two combs (as well as some broader noise artifacts) where upright and inverted triangles mark the two combs with distinct spacing in frequency.}
    \label{fig:common-lines}
\end{figure}

Glitches and lines can be further categorized by the morphology of the artifact. Many common glitch classes have been named (e.g., Tomte, Fast Scattering, Low Frequency Burst) based on a combination of their shape in spectrogram plots and what is known of their origin, see \fref{common-glitches}~\cite{Zevin:2016qwy, Glanzer_2023}. Some glitch classes are more detrimental to transient gravitational-wave searches and \ac{CBC} parameter estimation than others. Lines, by contrast, have fewer morphological distinctions, see~\fref{fig:common-lines}~\cite{PhysRevD.97.082002}. The primary distinction is between individual line artifacts and combs of lines. Combs arise when narrowband noise that couples into the gravitational-wave data is not purely sinusoidal, creating a series of spectral peaks with common frequency spacing. Combs are particularly problematic because one source can impact multiple narrow frequency bands at the same time.

This paper describes the activities performed to characterize the strain data measured by the \ac{LIGO} detectors, and investigations of instrumental and environmental noise between the end of O3 and end of O4a. 
In \sref{ins_invs}, we describe the instrumental investigations. In \sref{event_valid} we describe the activities to promptly validate gravitational-wave candidates, using tools to analyze the data-quality surrounding the event. In \sref{data_qual}, we describe the use  of data-quality products in searches of gravitational-waves of different kinds  (compact binary coalescences, un-modelled transients, continuous waves and stochastic background). In \sref{summ_conc}, we summarize the results, draw conclusions and present prospects for the near future. 

\section{Instrumental Investigations}\label{ins_invs}

Instrumental investigations carried out by the \ac{DetChar} group are crucial for understanding the impact of various noise sources on detector data quality~\cite{bergerAPL23}. The \ac{PEM} investigations are often carried out at the sites and require a strong co-ordination between the \ac{DetChar} group and the instrument scientists~\cite{o3pem}. These investigations rely on vibration, acoustic, and magnetic injections, which enable us to estimate environmental couplings between different parts of the detector and the \ac{GW} strain data. As explained in detail later in this section, some noise couplings can be reduced or eliminated, thereby reducing the amount of noise in \ac{GW} strain channel. In addition to artificially inducing environmental noise, we also routinely induce differential-arm displacements via the photon calibrators~\cite{Karki:2016pht} to study coupling between the \ac{GW} channel and interferometer auxiliary channels (see \sref{safety_section}) and add simulated gravitational waveforms for end-to-end testing of analysis pipelines~\cite{Biwer:2016oyg}.

In addition to artificially induced environmental noise and signals studies, the 
\ac{DetChar} group also analyzes instrument data in other ways to enhance our understanding of the detector. These studies usually use several \ac{DetChar} tools for determining the coupling between the environment, auxiliary sensors, and the detector noise characteristics. For example, through these investigations we may find that ground motion at a specific location in the detector is more correlated to the noise in the GW strain channel than at other locations. Such hypothesis can then be tested using the PEM tests. 


These investigations are central to solving problems that adversely impact the detector data-quality and uptime, consequently reducing the number of gravitational-wave observations. The transient noise impacts the parameter estimation process and could generate false alerts, which have to be retracted later. In this section, we first give an overview of transient noise in O4a at both the detectors, and then we discuss several \ac{DetChar} investigations carried out between the end of O3 and the end of O4a.

\subsection{Transient noise investigations at both sites}\label{both_invs}

\subsubsection{Transient Noise}\label{transients_both_sites}
\begin{figure}[h]
    \centering
    \includegraphics[width=0.7\textwidth]{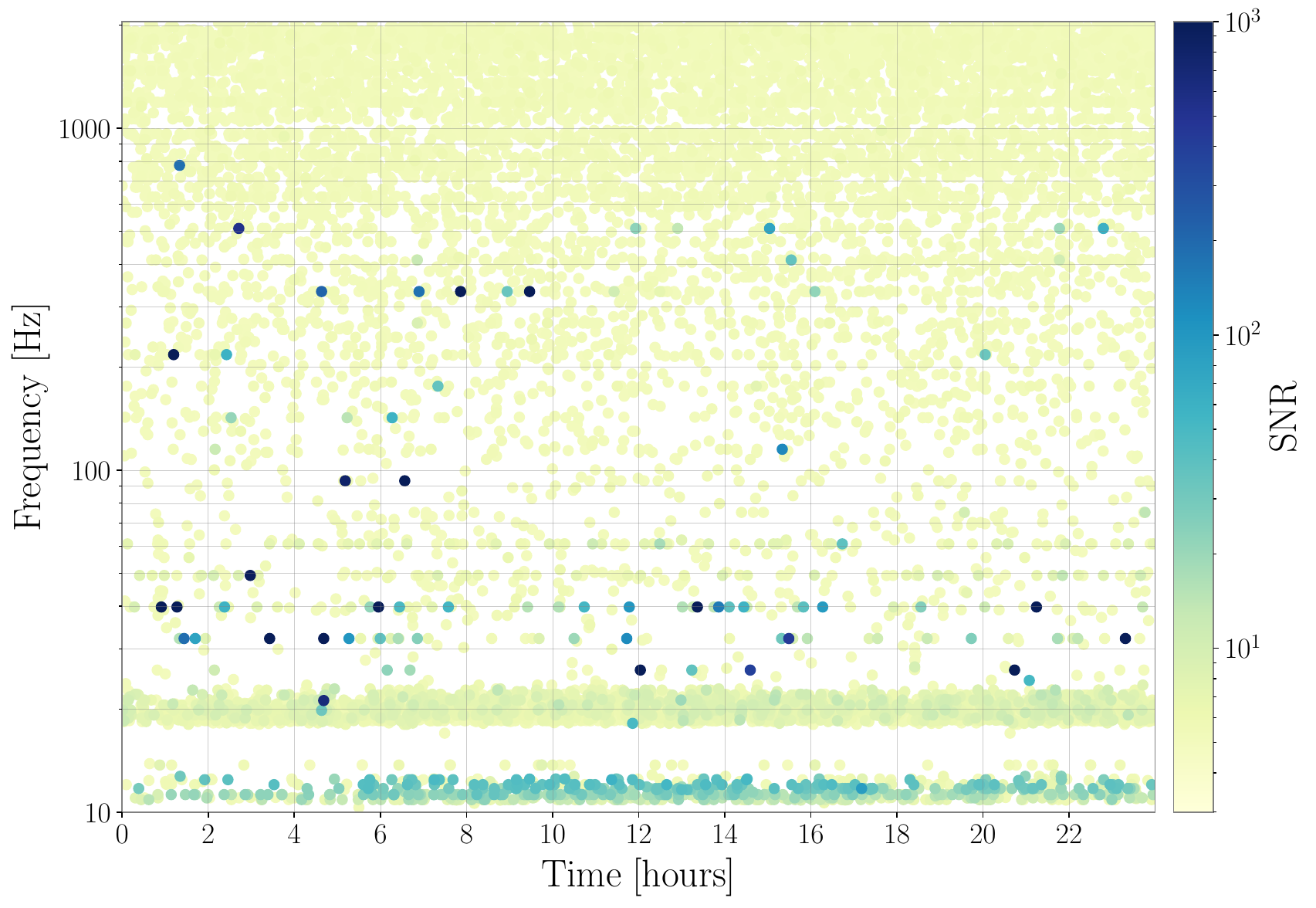}
    \caption{A glitchgram of Omicron triggers produced using \ac{LLO} data for one day during O4a; with its peak frequency plotted on the vertical axis and its \ac{SNR} indicated by the color.}\label{glitchgram}
\end{figure}

Glitches are short-duration bursts of excess power with their origins in environmental and/or instrumental couplings. Omicron is an event trigger generator used to search for this excess power in the primary gravitational channel, $h(t)$, and auxiliary channels \cite{Robinet:2020lbf}. The time, frequency, and \ac{SNR} of short-duration transient noise can be visualized using ``glitchgrams''. \Fref{glitchgram} shows an example of a glitchgram which is generated by plotting Omicron triggers.

Different tools are used to investigate glitches, and one of them is Gravity Spy, which classifies noise transients by their morphologies in spectrograms \cite{Zevin:2016qwy, Soni:2021cjy, Glanzer_2023, Zevin:2023rmt}. \Fref{common-glitches} shows time-frequency spectrograms for six common glitch categories at the two sites ~\cite{brown1991calculation,chatterji2004multiresolution}. Certain environmental or instrumental conditions can generate similar signals. Glitch classification enables the identification of patterns in the data for similar noise transients. 

Examining the relationship between noise in the primary gravitational-wave channel ($h(t)$) and multiple auxiliary channels can potentially lead us to the source of the noise. Hveto is one of the main tools used to study the correlation between transient noise in the primary strain and auxiliary data \cite{Smith:2011an}. Auxiliary channels that also witness the same noise are called witness channels. 
The glitchgram shown in \fref{glitchgram} is from a day when seismometers recorded elevated ground motion, which caused the excess transient noise at low frequencies around $10$--$20~\mathrm{Hz}$. Sometimes, this motion is so strong that the interferometer cannot maintain the servo-controlled resonance condition and stops observing~\cite{68196}. This is referred to as losing lock of the interferometer.

\begin{figure}[h]
    \centering
    \includegraphics[width=0.7\textwidth]{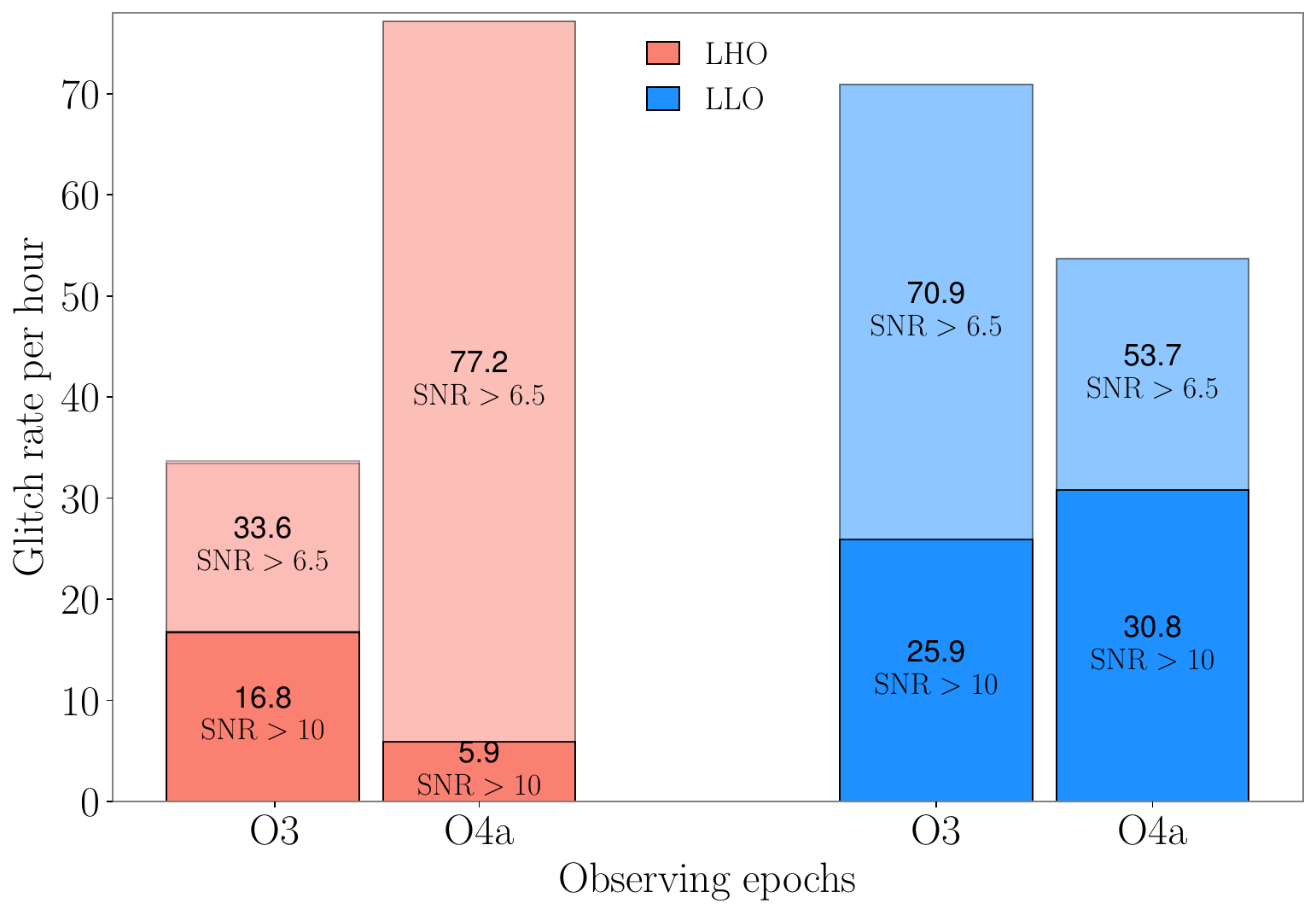}
    \caption{Glitch rates from Omicron during O3 and O4a at \ac{LLO} and \ac{LHO} in the frequency band $10$--$2048~\mathrm{Hz}$. For each observatory, two distinct rates were calculated: with \ac{SNR} above 6.5 and with \ac{SNR} above 10.}\label{glitch-rate}
\end{figure}

Multiple variables affect the glitch rate in the detectors, including instrumental upgrades, addition of new detector components, and environmental conditions such as wind, elevated ground motion, or the passing of trucks or trains near the site. \Fref{glitch-rate} shows the comparison of glitch rates between O3 and O4 for two different \ac{SNR} thresholds. In O3, transient noise at both detectors was dominated by stray light. These couplings were greatly reduced during O3b and after O3. Further details are discussed in  \cite{LIGO:2020zwl, Soni:2023kqq} and in \sref{llo_invs}.

\begin{figure}[h]
    \centering
    \includegraphics[width=\textwidth]{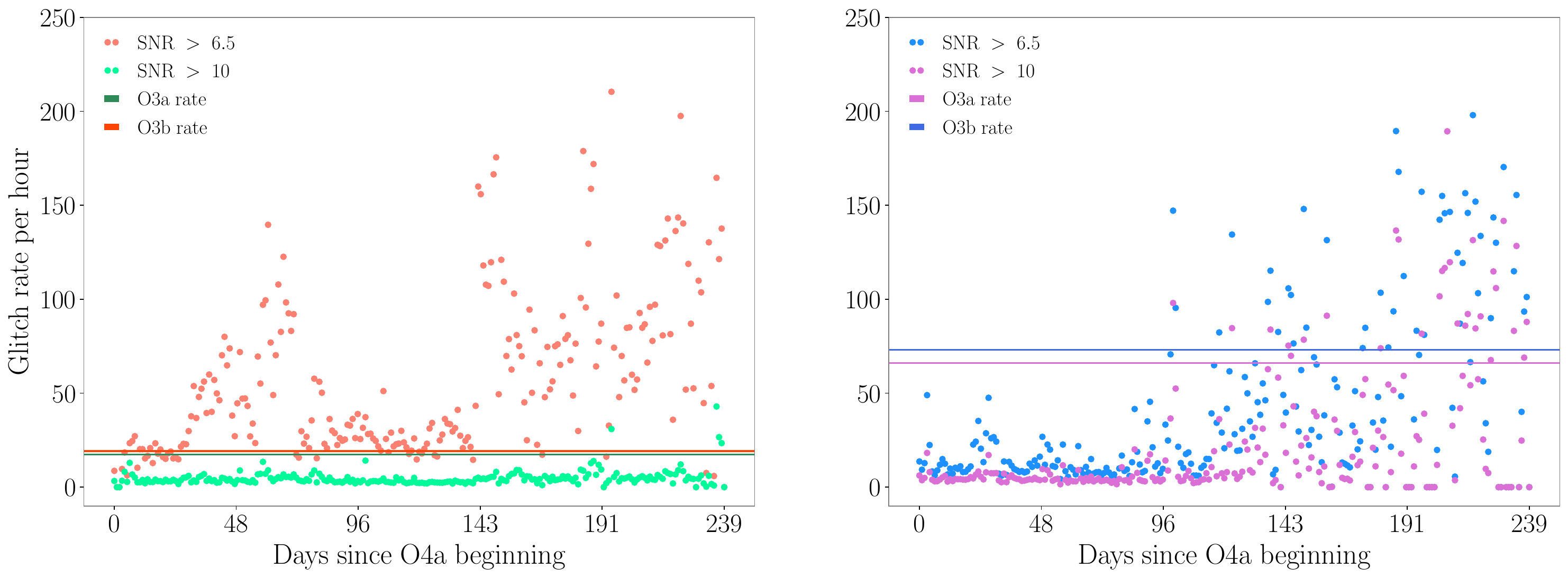}
    \caption{Omicron trigger rate during the O4a at LHO (left panel) and LLO (right panel) in the frequency band $10$--$2048~\mathrm{Hz}$. At \ac{LHO}, most of the glitches had \ac{SNR} $<$10. At \ac{LLO}, there is a visible increase in the rate of transients around the start of October 2023. This is due to unidentified stray light coupling induced by the seasonal increase in the microseismic ground motion.}
    \label{glitch_rate_fig}
\end{figure}

The O4a transient noise at LHO was dominated by low \ac{SNR} glitches mostly in $10$--$50~\mathrm{Hz}$. Most of the LHO transient noise was low \ac{SNR} as can be seen in \fref{glitch-rate} and in \fref{glitch_rate_fig}, which shows the daily glitch rate at LHO and LLO during O4a. These broadband transients had a common source and were mitigated during O4a. \Sref{low_snr_glitches} provides more details on this.

The transient noise at LLO was dominated by low-frequency ground motion that induce laser light scattering. Most of this ground motion can be attributed to the impact of atmosphere driven ocean waves on the ocean surface, also known as microseism \cite{microsiesm_paper}. The microseismic motion is seasonal and is caused by intense ocean wave activity from winter storms. This is why we see an increase in the glitch rate during the latter half of O4a as shown in \fref{glitch_rate_fig}. \Sref{slow_scatter_llo} provides more detail on this noise.

\subsubsection{Overview of scattering noise}
Noise due to light scattering is a common problem at both  detectors. Scattering happens when a small fraction of laser light gets reflected off some optics, hits another moving surface, and then rejoins the main laser beam. This rejoining leads to the introduction of a time-dependent phase modulation to the main laser beam~\cite{2012OExpr..20.8329O}. The additional phase noise shows up as $h_{\rm{ph}}(f)$,

\begin{equation}
  h_{\rm{ph}}(f) = \frac{K}{2}\frac{\lambda}{4\pi L}\mathcal{F}[\sin{\delta \phi_{sc}(t)}] \label{eq1}
\end{equation}
where

\begin{equation}
    \phi(t) = \phi_{0} + \delta \phi_{\rm{sc}}(t)  = \frac{4 \pi}{\lambda}[x_{0} + \delta x_{\rm{sc}}(t)]. \label{eq2}
\end{equation}
Here, $K$ is the ratio of stray light amplitude to the amplitude of light in the main beam (usually unknown but small, of the order of $10^{-9}$ in O4a microseismic scatter), $\mathcal{F}$ indicates a Fourier transform, $\lambda$ is laser wavelength (1064~nm) and $L$ is the length of interferometer arms (4~km), $x_{0}$ represents the static path length, $\delta x_{sc}(t)$ is the time-dependent motion of the scattering surface, which results into additional phase $\delta \phi_{sc}(t)$ on top of the static phase $\phi_{0}$ \cite{Accadia:2010zzb, Ottaway:12, LIGO:2020zwl, scatter-presentation}.
When the relative motion between the scattering surfaces is not small compared to the laser wavelength, we get fringe wrapping and the noise shows up as arches in time-frequency spectrograms (see \fref{common-glitches}). Transient scattering noise can be classified into two categories, which depends on the frequency of the ground motion that produces it: Slow Scattering and Fast Scattering \cite{LIGO:2020zwl}. Scattering is discussed in the following section, and again in \sref{llo_invs}.

\subsubsection{Cryo-manifold baffle noise}\label{cb_both_sites} Most of the optics in LIGO are housed in corner and end stations (X-arm and Y-arm).
During O3, vibrating \ac{CBs} at the \ac{LIGO} end stations were identified as a cause of light scattering noise~\cite{55927,Soni:2023kqq}. These baffles prevent most light reflecting from the beamtube reduction flange, where the beamtube narrows at the corner and end stations, from reentering the detector arms~\cite{dcc_stray_light_baffles_design}. Some light reflected from these baffles, however, still interferes with the circulating light in the main beam. When a mechanical resonance of the CB with a high quality factor was rung up from heightened ground motion, scattered light noise at about 4~Hz and harmonics were visible in $h(t)$.

At both detectors, rubber viton dampers were installed on three of the four \ac{CBs} to decrease the quality factor of the mechanical resonance prior to the start of O4a. Dampers were not installed on the final \ac{CB} because no other in-vacuum  maintenance work took place in its vicinity during the O3$-$O4a break. Damping reduced the velocity of the reflecting surface and thus lowered the cutoff frequency of the scattering noise to low frequencies where the interferometer is not sensitive. The vibration coupling at the remaining  undamped cryobaffles increased between O3 and O4 as the power in the arms increased. The difference in coupling between 60 W and 75 W input power at the undamped \ac{CBs} was about a factor of about 3 at LHO~\cite{70808}.

In the O4a-O4b commissioning break an incursion was made into each detector's X-arm end station vacuum enclosure to damp the final cryo-manifold baffle~\cite{69280}. After the viton damper installation, the quality factor of the $\sim$4~Hz mechanical mode in the CB dropped by a factor of $\sim$20, reducing the maximum frequency of the scattering noise by roughly the same amount~\cite{75553}.

\subsubsection{Safety studies}\label{safety_section}
A crucial aspect of verifying gravitational-wave candidates is ensuring that they are not introduced by environmental noise sources observed in auxiliary channels. In general, however, the transfer functions between $h(t)$ (GW strain) channel and the thousands of auxiliary channels is unknown, so it is possible for real gravitational-wave events to produce signals in some of the auxiliary channels. If we were to use such a channel (an ``unsafe'' channel) to veto a gravitational-wave event, without knowledge of this coupling, we would in effect be using an astrophysical signal to veto itself.
To probe the ``safety'' of auxiliary channels for vetoing candidate events, we use the photon calibrators to inject sine-Gaussian waveforms into $h(t)$ at each detector, mimicking a gravitational-wave signal. A safe auxiliary channel should not respond to these injected waveforms in $h(t)$. Then, we perform a statistical analysis of the $\mathcal{O}(5000)$ auxiliary channels sampled above 16~Hz using the \texttt{pointy-poisson} tool \citep{Essick:2020cyv} to classify auxiliary channels as either ``safe'' (i.e., acceptable to use to veto potential gravitational-wave events) or ``unsafe.'' The resulting list of safe channels is then passed on to downstream data-quality analyses. 

The injection and safety analysis process was repeated every few months in each detector during Engineering Run 15 (ER15) (April 26, 2023 - May 24, 2023) and O4a (May 24, 2023 - January 16, 2024) to track possible changes in safety and to ensure that any new channels were correctly classified as either safe or unsafe.
During the first set of analyses in ER15, a handful of \ac{LHO} channels were found to be unsafe compared to the existing list from O3: the \ac{ESD} voltage monitors at X-arm end station, and a handful of suspension rack magnetometers.

We found no substantial changes in channel safety during the subsequent duration of ER15 and O4a.

\subsubsection{\label{sec:spec-artifacts}Narrow spectral artifacts}
A typical daily-averaged, high-resolution spectrum ($\Delta f = 1/1800$~Hz) at either \ac{LHO} or \ac{LLO} reveals hundreds or thousands of narrow spectral artifacts (lines) within 10-2000~Hz, the band of particular interest to persistent gravitational-wave searches. These lines display a variety of amplitudes, widths, and shapes. Some are stable over long periods of time (weeks, months), while others are variable. Identifying and mitigating the most problematic lines requires both routine monitoring and focused investigations.

Since noise lines impact \ac{CW} and stochastic gravitational-wave searches in slightly different ways, specific tools are used to evaluate these artifacts to aid the analyses. \ac{CW}-focused line studies typically use a tool called \texttt{Fscan}~\cite{Fscan}, while stochastic-focused studies use tools called \texttt{STAMP-PEM}~\cite{Meyers:2018nyo} and \texttt{StochMon}~\cite{Stochmon}. These tools provide complementary information about lines, and their results are often used together to inform line investigations, mitigation efforts, and data-quality products.

\paragraph{Line investigations with Fscan}

\texttt{Fscan} produces high-resolution spectra averaged over long periods of time. It was largely rewritten for O4: modernizing the code, improving stability of data generation, making new data products and visualization tools available for analysis, and enabling production of custom spectra. In O4a, \texttt{Fscan} was used to generate daily, weekly, and monthly spectra for about 80 channels at each observatory site, using \acp{FFT} of 1800-s-long data segments. Additional analyses were performed to track lines of interest (determining witness channels and times at which the artifacts changed) using \texttt{Fscan} data.

In \sref{lho-combs}, we highlight examples of successful investigation and mitigation efforts in O4a at LHO. Because LLO has generally cleaner data for persistent gravitational-wave searches, there have been fewer notable examples of mitigation. A number of high-priority narrow spectral noise artifacts, however, have not yet been mitigated, including artifacts present at both detector sites. Additional work is ongoing in this area. The highest-priority artifacts are those that contaminate a broad spectral region (i.e., combs, especially those with many visible peaks) and artifacts that are present at both sites (e.g., 60 Hz power mains) because these have a disproportionate impact on persistent gravitational-wave searches.

The highest line artifacts in \fref{O3bO4aASDs_fig} have known cause, typically due to choices inherent to the detector design (mirror suspension resonances at various frequencies, strongest at 300~Hz and 500~Hz and harmonics) or calibration and dither lines to monitor or control interferometric cavities, respectively.

\paragraph{Monitoring strain-strain narrow-band coherence.} 

\begin{figure}
    \centering
    \includegraphics[width=\textwidth]{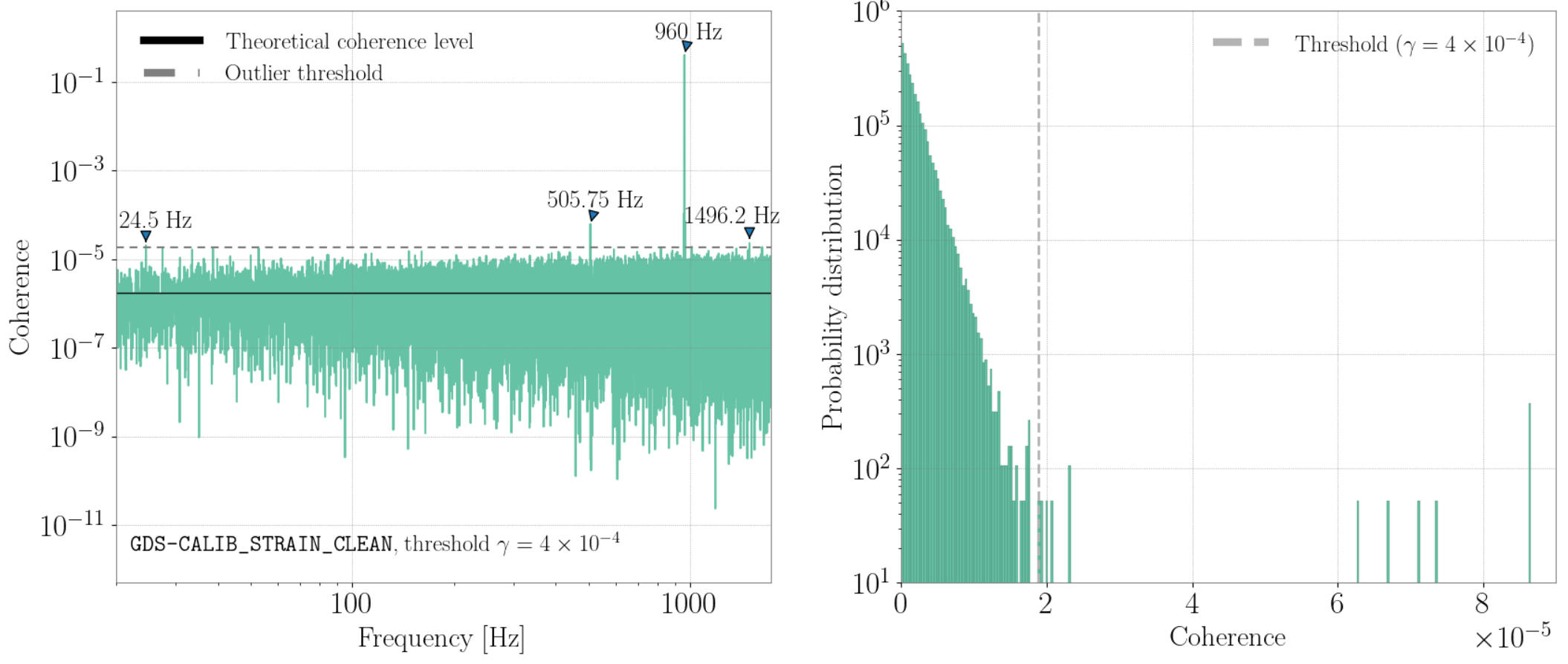}
    \caption{Left panel: coherence spectrum between \ac{LHO}-\ac{LLO} interferometer strain readouts. Right panel: histogram of the spectrum from \texttt{StochMon}.}
    \label{fig:coherence_stochmon}
\end{figure}

Stochastic searches rely on cross-correlating $h(t)$ data from detector pairs, so understanding and monitoring potential noise sources that could detrimentally impact the cross-correlated data is crucial. A stochastic monitoring tool called \texttt{StochMon} is specifically designed to calculate strain--strain coherence in medium-latency (daily pages produced within 24 hours) and flag problematic frequency bins that have excess coherence. %
The expected random coherence between Gaussian datastreams is defined as the inverse of the effective number of segments averaged to produce the coherence spectrum, $1/N_{\rm eff}$; in the case of \texttt{StochMon}, Welch averaging with a 50\% overlap is used between consecutive segments (for details on the $N_{\rm eff}$ calculation see App. A of~\cite{pygwb}). %
Frequency bins are flagged when they exceed a Gaussian coherence threshold of 
\begin{equation}
\gamma = 1 - \left(\frac{1}{N_{f}}\right)^{1/(N_{\rm eff} -1)}\,,
\end{equation}
where $N_f$ is the number of frequency bins. %
An example of the calculated coherence and its outliers is shown in \fref{fig:coherence_stochmon}. The outlier bins shown have all been traced back to specific detector noise sources. The $505.75~\mathrm{Hz}$ and $1496.2~\mathrm{Hz}$ outliers lies within to the first and third violin modes of the detectors, which are resonances of the detector’s mirror suspension fibers and are caused by changes in the fiber tension. The $24.5 ~\mathrm{Hz}$ outlier corresponds to a calibration line that was turned on between July 25, 2023 to August 9, 2023 ~\cite{72096}. The $960 ~\mathrm{Hz}$ outlier is caused by one of the DuoTone signals in both LIGO detectors from the timing system, which is used to synchronize data collection across the global detector network and within individual interferometers. ~\cite{64525}. 

In practice, the investigation to determine the instrumental causes of the coherence outliers is crucial. They are carried out by spectral monitoring tools including \texttt{Fscan}, and \texttt{STAMP-PEM}. \texttt{STAMP-PEM} is another stochastic monitoring tool that keeps track of daily and weekly coherence between $h(t)$ and physical environmental channels at 0.1 Hz resolution.  In terms of computing resources, a moderately low resolution of 0.1 Hz allows it to monitor a wide range of channels (about 1000 channels per observatory site). The high-resolution spectral information from \texttt{Fscan} and auxiliary channel information from \texttt{STAMP-PEM} provide complementary information to support the strain-strain coherence outliers investigation.


\subsubsection{Broadband persistent artifacts}

\begin{figure}
    \centering
    \includegraphics[width=0.9\textwidth]{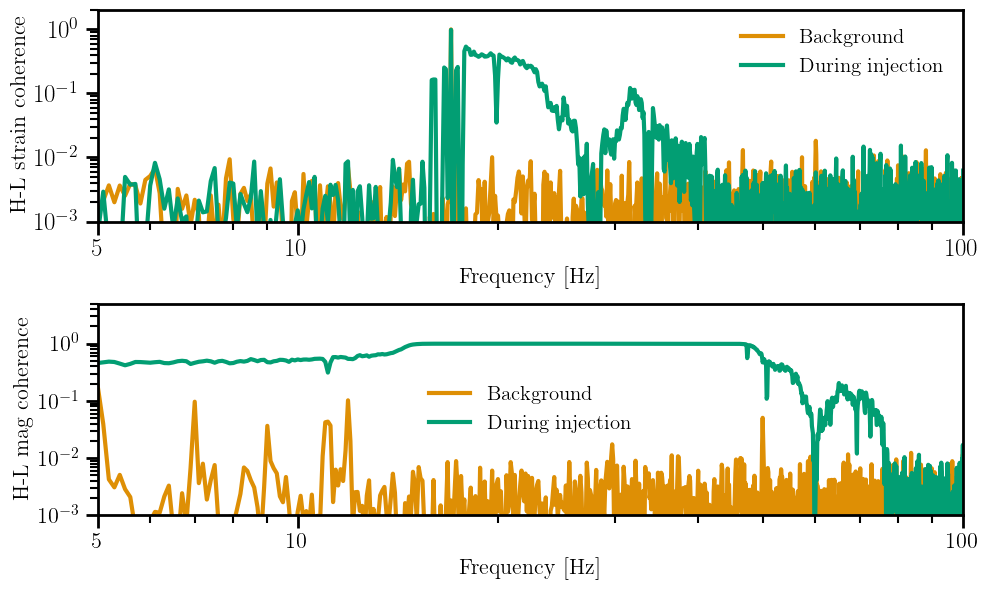}
    \caption{Coherence spectrums between \ac{LHO}-\ac{LLO} $h(t)$ (top panel) and \ac{LHO}-\ac{LLO} magnetometers (bottom panel) during the correlated magnetic injections with Schumann-like spectrum, and the reference background time taken on December 21, 2023.}
    \label{fig:stochastic correlated injections}
\end{figure}

\paragraph{Investigations of coherence noise}  
Correlated magnetic noise was investigated as a potential noise source for stochastic searches. During O4a, two sets of coordinated magnetic injections between the two LIGO interferometers were performed to study the coupling between the coherence of $h(t)$ in the presence of an increased correlated magnetic field, thereby determining the magnetic noise budget. The first set of injections lasted 5 minutes and was composed of broadband white noise to test the synchronous injections capability between the sites. The second set lasted approximately 45 minutes and consisted of a spectrum with Schumann-like frequency characteristics but at a significantly higher intensity ($\sim100 ~\mathrm{pT}$  in comparison with a realistic amplitude of Schumann resonance magnetic field of $\sim1 ~\mathrm{pT} $ ) (see e.g.,~\cite{Janssens:2021cta} for a description of Schumann noise relevant for ground-based gravitational-wave detectors). \Fref{fig:stochastic correlated injections} shows the coherence spectra between the x direction magnetometers in Laser and Vacuum Experimental Area in the corner station as well as the coherence between  $h(t)$ at LHO and LLO during the second set of injections compared to a reference background time. Details on the level of correlated magnetic noise estimate will be provided in the O4 stochastic analyses release.

\subsection{LIGO Hanford noise investigations}\label{lho_invs}
\subsubsection{Electronics ground noise}
During the commissioning period before O4, spectra of the variation in current flowing from building electronics ground to neutral earth were observed to be correlated with noise in $h(t)$.  A newly developed electronics ground injection system showed that noise in $h(t)$ could be produced by injecting $\sim$100~mA currents onto the building electronics ground. The coupling is thought to be produced by fluctuations in the potential of the electronics ground system due to the variations in current flows across the finite resistance between electronics ground and true neutral earth, measured to be about 2 $\Omega$ at \ac{LHO}~\cite{67075}. Forces on the charged test mass may fluctuate with the potentials of nearby electronic systems that are referenced to the fluctuating electronics ground, such as the electrostatic drives (ESDs) and ring heaters. 

The noise from electronics ground potential fluctuations was reduced in two ways. First, the resistance between certain electronics chassis and the building electronics ground were reduced in order to reduce the total resistance to neutral earth for those electronics. At LHO it was found that lowering the resistance on the grounding wires for electronics chassis used to control test mass motion made the electronics less sensitive to ground potential fluctuations~\cite{65907, 66469}. Changing the grounding of controls chassis located at the end stations lowered the noise in $h(t)$ overall and also reduced coherence between test mass motion and current to ground below 100~Hz~\cite{67075}.

Second, the biases of the \ac{ESD}s were swept and set to values that minimized the coupling to $h(t)$ of injections onto the electronics ground. It is thought that, at the coupling minimum, the forces on the charged test mass due to ground potential fluctuations are partially canceled out by an opposite dipole force associated with the bias-polarization of the test mass~\cite{67075}. Sensitivity to ground potential fluctuations was therefore further reduced by selecting a bias voltage for the DC component of the \ac{ESD}s which minimized coupling between $h(t)$ and currents injected onto an electronics chassis at each end station~\cite{67075}.

\begin{figure}
\centering
\includegraphics[width=\textwidth]{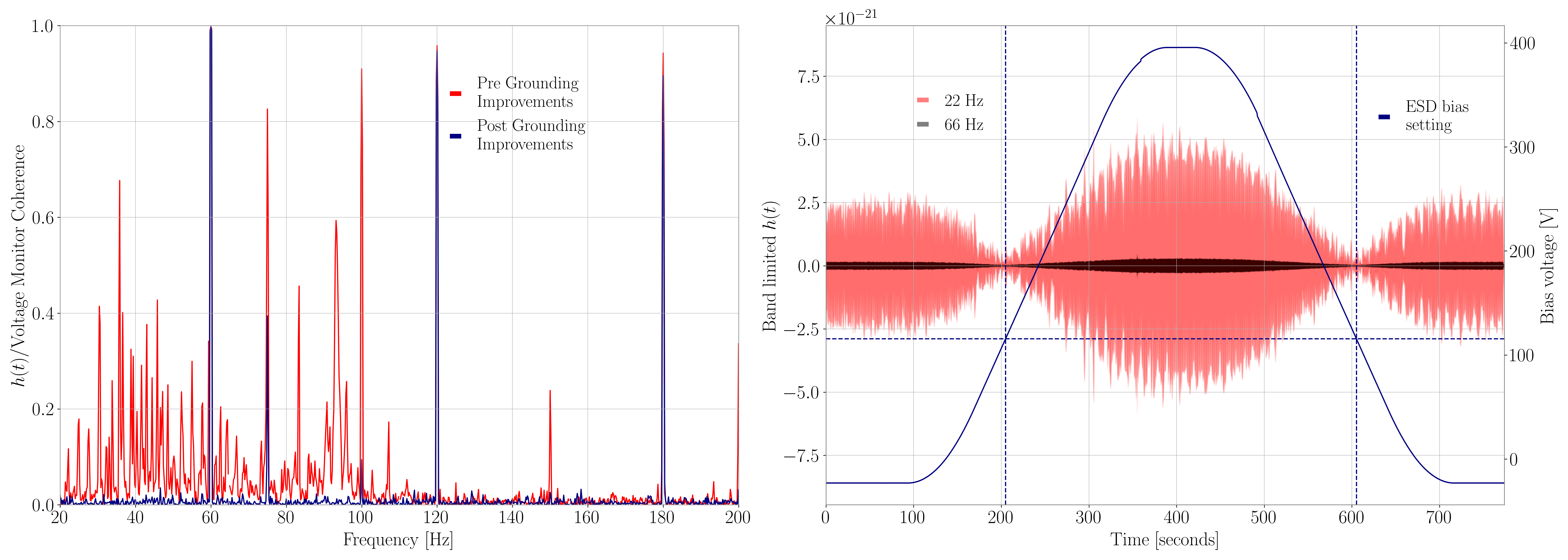}
\caption{\label{fig:elec_ii} Two methods for reducing electronics noise in the \ac{GW} data. Left panel: The coherence between the $h(t)$ data and a temporary voltage monitor installed on a controls chassis at the X-arm end station~\cite{66469}. Coherence between $h(t)$ and  chassis voltage decreased after changing the grounding wires used on controls chassis. Right panel: the bias voltage supplied to the \acp{ESD} is modulated, affecting the coupling of the harmonics of an 11~Hz comb injection. The new bias voltage was subsequently set to the value which minimized the amplitude of $h(t)$ around the 11~Hz harmonics during the test~\cite{67075}. Dashed lines show that that an \ac{ESD} bias voltage of $\sim$115~V minimizes noise in $h(t)$.}
\end{figure}

These mitigations resulted in \ac{BNS} range improvements of a few megaparsecs. The effects of these two changes on $h(t)$ at LHO is illustrated in \fref{fig:elec_ii}. Further mitigation could be obtained by shielding electronics inside the chamber from the test mass with shields that are connected to the chamber so that charges can rearrange to cancel the low-frequency fields produced by the electronics. The bias voltage that minimizes currents coupling to $h(t)$ changes over time and continues to be tracked~\cite{72118}.

A minimum noise setting for Y-end \ac{ESD} bias was identified at LLO before O4~\cite{64609} and midrun changes in the X-arm end station-end \ac{ESD} bias were found to reduce noise at $\sim$11~Hz and $\sim$60~Hz and harmonics of these frequencies~\cite{67581}.

\subsubsection{Broadband transient noise}\label{low_snr_glitches}

We noticed an increased noise at low frequencies which showed non-stationary behavior in the frequency band $10-50~\mathrm{Hz}$. A bicoherence analysis of  $h(t)$ noise with itself found that this noise was modulated by a low frequency $h(t)$ signal mostly around 2.6 Hz \cite{71005, 71092}. Most of the longitudinal drive control to the ESD was being sent in the band $1-3~\mathrm{Hz}$ which could have contributed to this increased noise. A new longitudinal control scheme that reduced the amount of control sent at these low frequencies was developed and implemented . This led to significant improvement in $h(t)$ noise, thereby reducing the non-stationary or glitch behavior as well~\cite{71927}.

\subsubsection{Modifying input power}
The amount of laser power input into the \ac{LIGO} interferometers has increased in each Observing run. Increased power circulating in the arms improves the high-frequency sensitivity of the \ac{LIGO} detectors by reducing the effect of quantum shot noise. In O4a, both \ac{LIGO} detectors were slated to operate with $75~\mathrm{W}$ of laser power sent into the \ac{IMC}. 

As input laser power was increased at both detectors prior to the fourth Observing run, vibration coupling also increased. This includes vibration coupling through both scattered light noise and input beam jitter noise. A possible explanation is that increased thermal distortion of the test mass surfaces around coating defects may increase scattered light and also reduces the symmetry of the arms, decreasing common mode rejection of input noise. The dramatic increases in coupling suggests that vibration coupling may become increasingly problematic as input power is increased~\cite{70808} in future Observing runs and next-generation observatories.

\begin{figure}
\centering
\includegraphics[width=\textwidth]{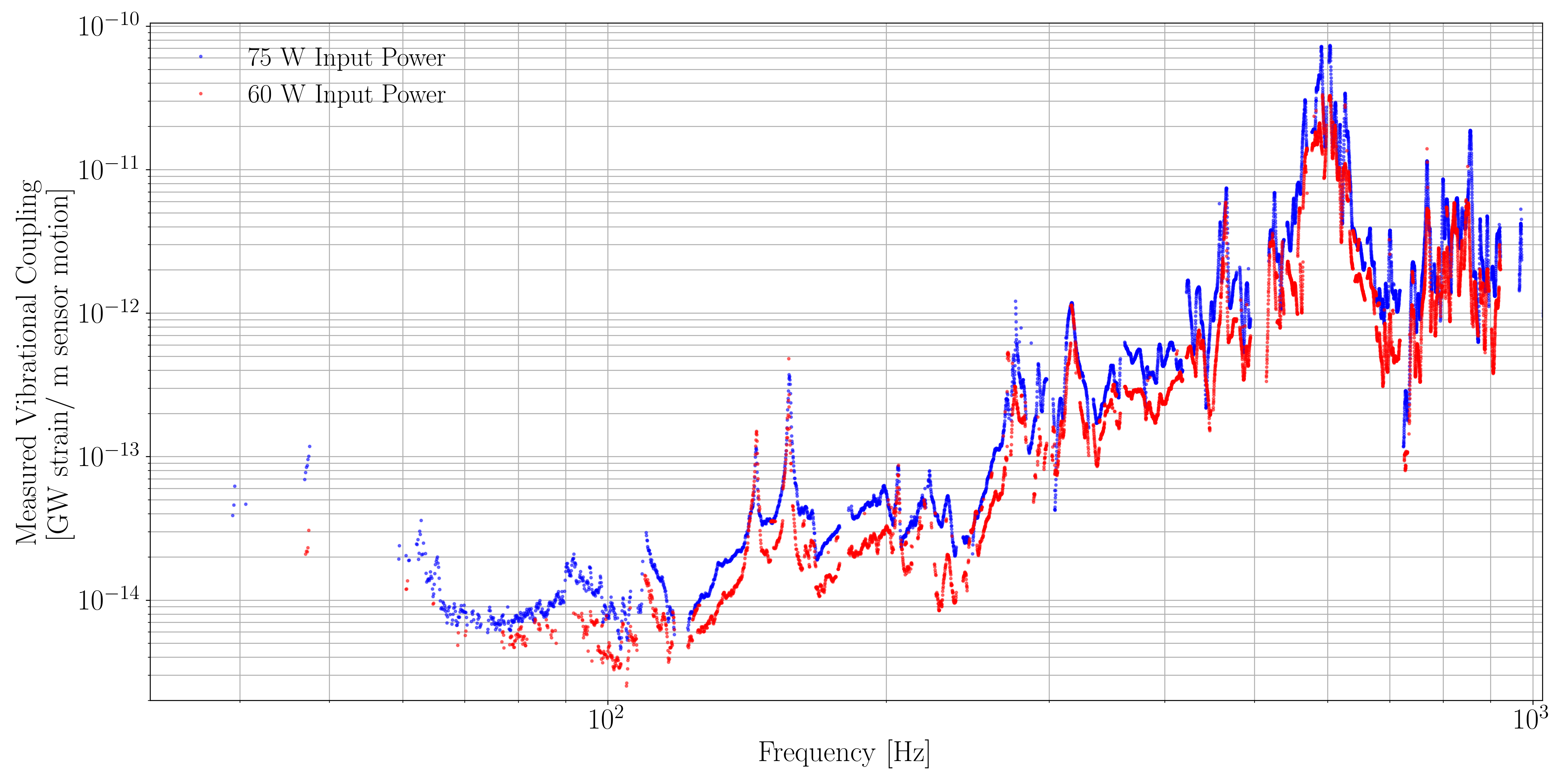}
\caption{\label{fig:pslccf} Comparison of the composite vibrational coupling function between \ac{PEM} sensors placed around the pre-stabilized laser optics table measured during $75~\mathrm{W}$ (blue) and $60~\mathrm{W}$ (red) operation. The measured vibrational coupling at the input laser, observed where both the \ac{GW} data channel and \ac{PEM} channels witness a \ac{PEM} injection~\cite{o3pem}, is worse at nearly all frequencies when operating at $75~\mathrm{W}$ of input laser power.}
\end{figure}

Due to duty cycle and control scheme concerns associated with high-power operation the laser power sent to the \ac{IMC} was reduced from $75~\mathrm{W}$ to $60~\mathrm{W}$ at LHO during O4a~\cite{70648}. \Fref{fig:pslccf} shows the reduction in vibrational coupling between \ac{PEM} sensors placed around the in-air optics table where the input laser light is produced and the apparent differential arm length.

\subsubsection{Weekly magnetic monitoring}

\begin{figure}
    \centering
    \includegraphics[width=\textwidth]{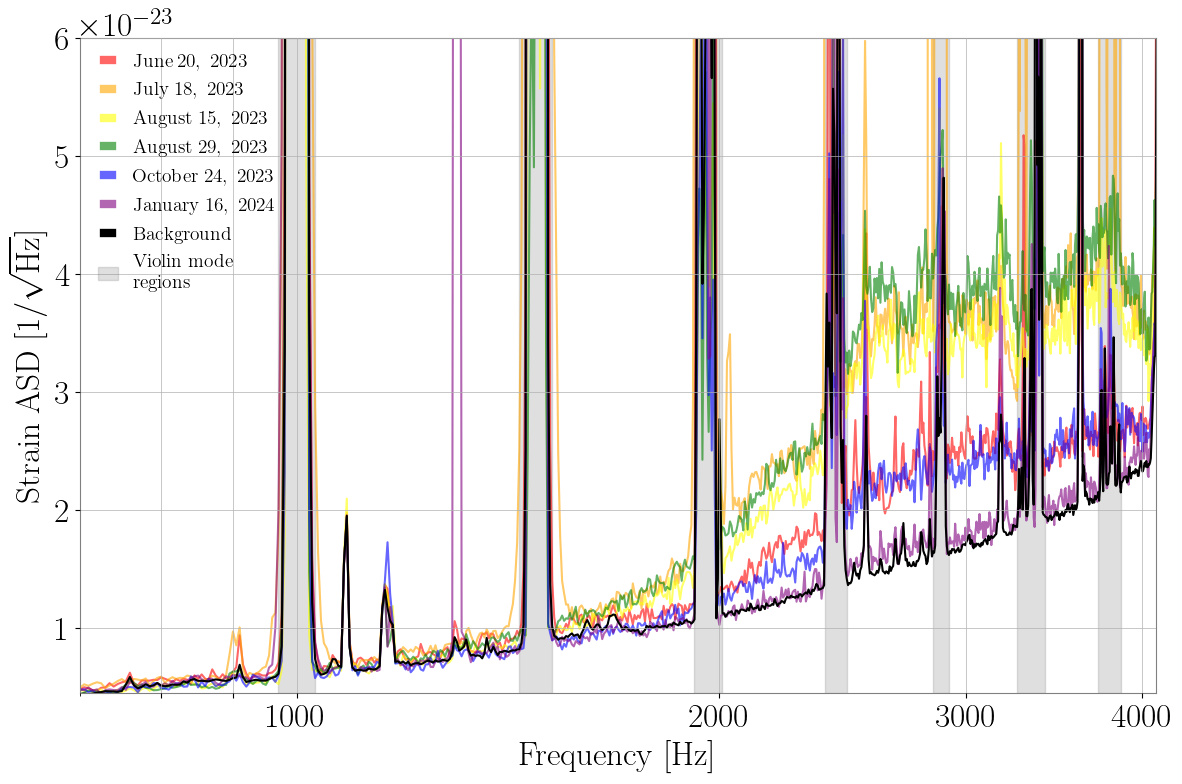}
    \caption{\label{fig:weeklymags} $h(t)$ data from 6 of the 22 weekly high-frequency magnetic injections performed at the \ac{LHO} corner station. Above $\sim$2000~Hz, there was a significant response in $h(t)$ when a large external magnetic field was applied near the beamsplitter area during the middle portion of O4a. The reference background time was taken on June 20, 2023. The strain ASD also varied from week to week in certain frequency bands due to the relative absence or presence of excited thermal (violin) modes in the data. The violin mode frequency bands are denoted by the grey regions in the figure.}
\end{figure}

Regular measurements were taken before and during O4a to understand the potential for coupling between local magnetic fields and the interferometer~\cite{64609,69745}. Local magnetic fields were generated by running a current through large coils of wire mounted in the experiment hall and near electronics racks used for interferometer controls~\cite{o3pem}. The response of each interferometer to the resulting magnetic fields was quantified using the network of PEM magnetometers set up around each observatory~\cite{s6pem,o3pem}. In order to vet GW candidates at kilohertz frequencies, such as transients from neutron star f-modes~\cite{fmodediscovery,fmodes,fmodedetection}, more computing space was allocated prior to O4a to store accelerometer and magnetometer data up to $8192$~Hz. Nearly every week of O4a, a broadband magnetic field was injected at $1000-4096$~Hz at $7$ locations around LHO for $36$~s at each coil to quantify magnetic coupling in the newly-monitored  part of the kilohertz regime. These injections provoked a response in $h(t)$ at the \ac{LIGO} LHO corner station. \Fref{fig:weeklymags} shows  $h(t)$ response to several of these weekly injections compared to a reference background time. Weekly probes of the high frequency magnetic coupling could be used to more accurately estimate the total environmental contribution to a high-frequency GW candidate~\cite{Helmling-Cornell:2023wqe}. Broadband magnetic injections were also made over $10-100$~Hz and $100-1000$~Hz from these $7$ coils as part of the weekly injection campaign.

During the course of O4a, the magnetic coupling of the detector at kilohertz frequencies fluctuated from week to week. At the beginning of O4a, the LHO detector was somewhat sensitive to large external magnetic fields applied at the Corner Station. Towards the middle of O4a, the detector's response to these applied magnetic fields increased, before dropping to being only weakly coupled towards the end of O4a. The most likely mechanism for the observed magnetic coupling in the kilohertz regime is magnetic interference with cables that control the \ac{LIGO} suspensions and optics. Specific mid-run electronics configuration changes which affected the degree of magnetic coupling, such as cables being rerouted, have not yet been identified.

\subsubsection{Cosmic ray glitch correlations}
A high energy cosmic ray shower is a potential mechanism for creating glitches via momentum transfer to, heating of, or changing the electric potential near the test masses~\cite{cosmicray_brag,cosmicray_yama}. Cosmic rays are monitored by four photomultiplier tubes placed beneath the vacuum chamber which houses the X-arm input test mass at LHO. During O2 and O3, we compared the time difference between cosmic ray arrival times and blip glitches at LHO and found no evidence of a correlation~\cite{o2o3_detchar}. We expanded this search in O4a to include blips, low-frequency blips, repeating blips, and tomtes as identified by Gravity Spy~\cite{Zevin:2016qwy}. A description of the cosmic ray sensor systematics in O4a can be found at~\cite{cr_systematics}. No temporal correlation was found between cosmic rays and any of these glitch classes. Additionally, we found the amplitude of cosmic rays which struck LHO within a second of a glitch were consistent with the overall amplitude distribution of cosmic rays witnessed by the cosmic ray detectors installed at LHO.

\subsubsection{Scattering noise from the input arm}
In O4a, short shutdowns of the LHO building \ac{HVAC} system produced several percent increases in astrophysical range~\cite{73941}. Localized vibration injections indicated that the coupling of the \ac{HVAC} vibrations at the corner station was in the input arm of the interferometer~\cite{74175}. The coupling was further localized by using broad-band shaker injections to increase the amplitude of vibrations above ambient levels so that laser vibrometry could be used to identify the internal structures that had resonances that were characteristic of the scattering noise~\cite{74772}. During the O4a-O4b break, the baffles were moved and damped, and new baffles added, greatly reducing the vibration coupling in the input arm~\cite{75726,76969}.

\subsubsection{\label{lho-combs}Comb investigations }

During O4a, two sources of comb artifacts were identified and mitigated at LHO. The first source was in the electronics driving a mirror heating element. This created a comb of approximately 1.6611~Hz (though different mitigation efforts caused changes in the spacing) centered around 280~Hz. A time-correlation was found between changing electronic settings for the mirror heating element and variations in the 1.6611~Hz comb amplitude, which suggested a possible source for the comb~\cite{71801}. This insight motivated subsequent mitigation efforts that more clearly identified the problem. Electrical connections were changed to stop the 1.6611~Hz comb from being created~\cite{73886}.

The second source created a near-1~Hz comb, as well as a near-5~Hz and near-7~Hz comb at various times. All three combs were traced to Hartmann wavefront sensors (HWS), which are part of the interferometer \ac{ASC} subsystem~\cite{Barsotti_2010}. It was determined that the comb frequency spacing changed when the HWS camera shutter frequency setting was changed~\cite{74614, 74617}. The low amplitude of the comb makes observing this artifact in short stretches of data (less than $\sim$1 day) challenging, and thus mitigation efforts more time-consuming. Once the connection between changes in the comb spacing to changes in HWS hardware settings was established, efforts to change the hardware configuration while in observing mode helped to mitigate this comb~\cite{75876}.

\subsection{LIGO Livingston noise investigations}\label{llo_invs}

\subsubsection{Slow scattering}\label{slow_scatter_llo}
\begin{figure}[t]
    \centering
    \includegraphics[width=\textwidth]{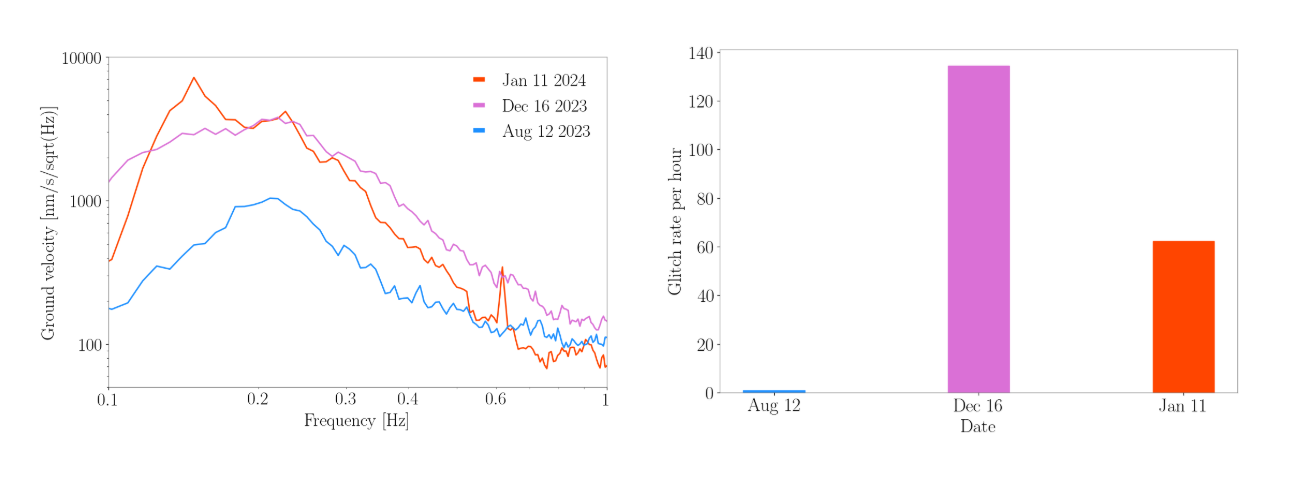}
    \caption{Left panel: Ground motion in the $0.1$--$1~\mathrm{Hz}$ frequency band during different days of O4a. Right panel: Hourly rate of glitches in the frequency band $10$--$60~\mathrm{Hz}$ and SNR in between $20$ and $200$ during the same days.}
    \label{fig:gm_and_glitch_rate}
\end{figure}

Noise due to high ground motion in the band $0.1$--$0.5~\mathrm{Hz}$ was the most dominant source of glitches in the \ac{LLO} data during O4a. These glitches, also known as Slow Scattering, adversely impacted the strain sensitivity mostly in $10$--$50~\mathrm{Hz}$ band. \Fref{fig:gm_and_glitch_rate} shows glitch rate and ground motion for three days. For two of these days, ground motion in the band $0.1$--$0.5~\mathrm{Hz}$ was high, which led to a high rate of Slow Scattering glitches in the data.
 
The additional phase noise as given by \eref{eq2} shows up as arches in the time-frequency spectrogram as shown in the middle plots in \fref{common-glitches}. The time separation between subsequent scattering arches gives a direct measure of the frequency with which the scattering surface is moving. During O4a, we have seen that the frequency of the scattering surface is not constant because it moves at whatever frequency is dominant in the ground motion~\cite{scatter-presentation}. We have not yet found any optics which have enough velocity to create noise above $10~\mathrm{Hz}$ in $h(t)$.

\subsubsection{Fast scattering}
During \ac{O3}, Fast Scattering glitches were the most common glitch source at \ac{LLO}, making up about 27\% of all glitches with a confidence of 90\%, according to Gravity Spy~\cite{Zevin:2016qwy,Soni:2021cjy, Glanzer_2023}. Fast scattering is typically found to be correlated with ground motion in the microseismic band $0.1$--$0.3~\mathrm{Hz}$, and the anthropogenic band $1$--$6~\mathrm{Hz}$. Fast scattering arches are short in duration, shown in \fref{common-glitches}, and impact the detector sensitivity in the $10$--$100~\mathrm{Hz}$ frequency range. Trains, logging, construction, and other human activity were the main sources of Fast Scattering, as the anthropogenic motion upconverts to higher frequency ~\cite{Glanzer_2023, Soni:2023kqq}.

In O3, trains near the \ac{LLO} Y-arm end station produced low-frequency seismic noise that would upconvert into the gravitational-wave sensitive frequency band \cite{Glanzer:2023hzf}. For this reason, they provided an avenue to study how periods of large ground motion impacted the detector. Spectrograms of the ground motion revealed many harmonic lines with changing frequency, and short bursts of increased amplitude in the strain data. The suspicion was that each burst was produced by the low-frequency ground motion exciting mechanical resonances of some scattering surface. Two methods, Lasso regression~\cite{Tibshirani-Lasso} and Spearman correlation~\cite{boslaugh-Spearman}, were employed to identify which narrow band seismic frequencies contributed the most to increased detector noise. Both methods consistently pointed to ground motion in the $1.8$--$2.2~\mathrm{Hz}$ range as the primary factor correlating with heightened strain noise at the corner station~\cite{Glanzer:2023hzf}. The subsequent mitigation of noise from these frequencies for O4 is discussed in \sref{acbres}.

\begin{figure}[t]
    \centering
    \includegraphics[width=\textwidth]{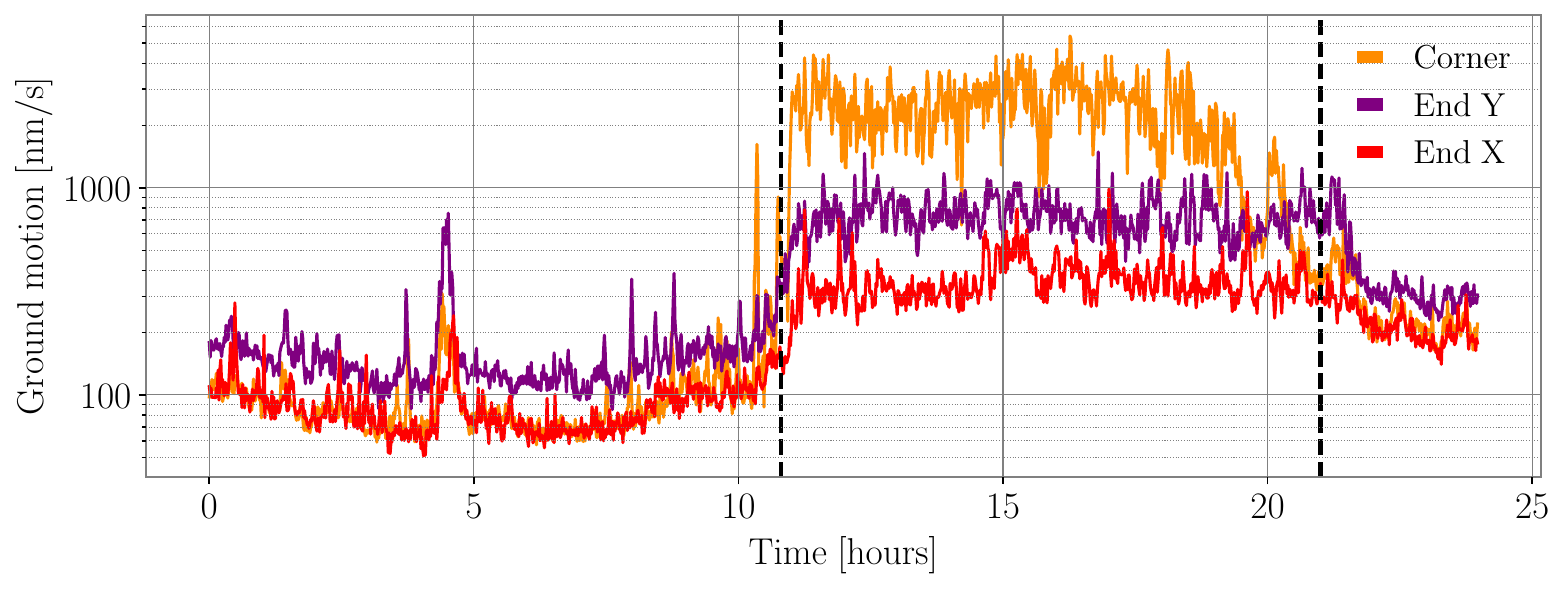}
    \caption{Example of a day in which \ac{LLO} lost lock due to nearby logging activities. Each trace represents  band-limited RMS data from $1$--$3~\mathrm{Hz}$ taken from seismometers at the X-arm end station, Y-arm end station and corner station. The detector could not be locked for roughly 10 hours, indicated by the dashed lines, as ground motion levels in the vertical direction surpassed 1000~nm/s at the corner station. In local central time, the elevated ground motion was from roughly 6am to 4pm. }\label{logging}
\end{figure}

From roughly June 2023 through August 2023, there was a significant amount of logging occurring near \ac{LLO}~\cite{65149,65404}. The anthropogenic ground motion in the vertical direction near the corner station consistently reached amplitudes greater than 1000~nm/s, as shown in \fref{logging}. These high ground motion levels caused the detector to lose lock multiple times for many hours during the daytime. After August 2023, the logging activity ceased, significantly reducing the disruptive ground motion near the corner station.

\subsubsection{Arm cavity baffle resonances}\label{acbres}
\begin{figure}[t]
    \centering
    \includegraphics[width=\textwidth]{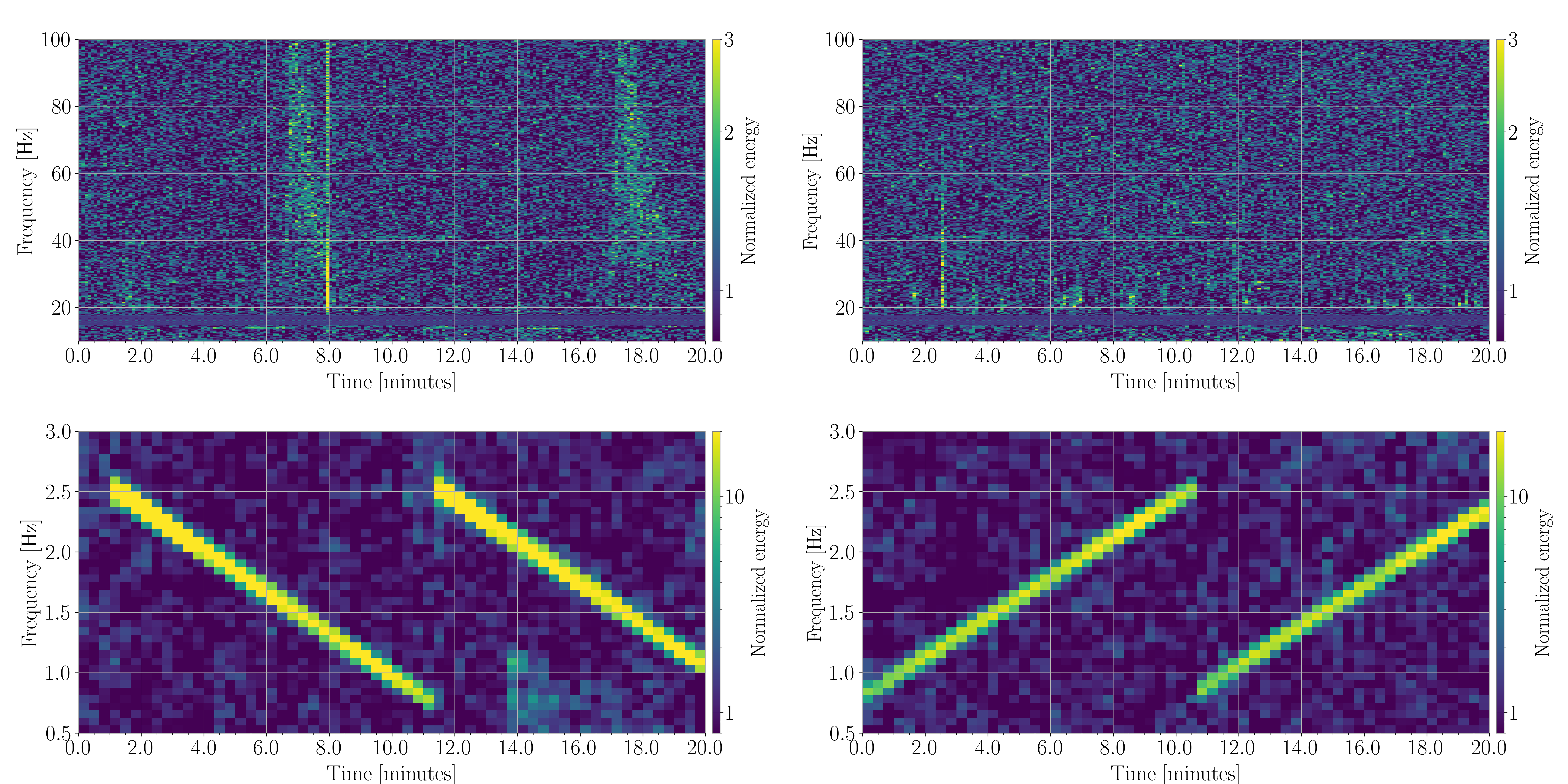}
    \caption{Sweep test comparison before and after the damping of the ACB~\cite{63699}. The top spectrograms show the strain noise, and the bottom spectrograms show the sweep injections performed on the corner station HEPI. After the ACB was damped, we no longer observed increased strain noise due to mechanical resonances.}\label{acb-sweep-test}
\end{figure}

Arm cavity baffles (ACBs)~\cite{dcc_ACB_design} are located at each of the test masses, attached to the first stage of the active seismic isolation system (\ac{HEPI}~\cite{Wen:2013boa}), and are used to catch the light from wide angle scattering. \ac{ACB} resonances are sensitive to the physical state of the system and changes to it can led to the shift in resonant frequencies~\cite{Soni:2023kqq}. After O3, but before O4, at the corner and Y-arm end station, the \acp{ACB} had a high-quality-factor resonance at around 1.6~Hz~\cite{60927}. When rung up, noise appears in the gravitational-wave data from around $20$--$100~\mathrm{Hz}$. In the absence of high microseismic ground motion, 1.6~Hz motion would create scattering noise at 3.2~Hz. After O3 ended, 3.3~Hz scattering noise was observed that had not been seen before; this can be explained by \ac{ACB} resonances at 1.6~Hz. 
We suspect that during O3, the \ac{ACB} resonance was around 2~Hz, which would produce the common 4~Hz Fast Scattering observed. In late 2022, the \ac{ACB} resonances at the corner and Y-arm end station were mechanically damped. As a result, the rate of Fast Scattering decreased dramatically and subsequently it was found that this noise coupling was no longer present \cite{Soni:2023kqq}. The effect of this remediation can be seen in the sweep injections performed in July 2022 (see \fref{acb-sweep-test}) using a shaker and again in February 2023~\cite{63569}.

Comparisons of the impact of logging during O3 and O4 showed that for similar ground motion amplitudes, the rate of transients was significantly reduced by a factor of about 50~\cite{Soni:2023kqq}. In O3, anthropogenic ground motion at such high levels would have produced many glitches detected by Omicron (assuming lock is not lost). Due to the damping of the arm cavity baffles, we did not observe significant strain noise due to the logging activities in O4~\cite{65494}. 

\subsubsection{Binary neutron star range oscillations}
\begin{figure}
    \centering
    \includegraphics[width=0.7\textwidth]{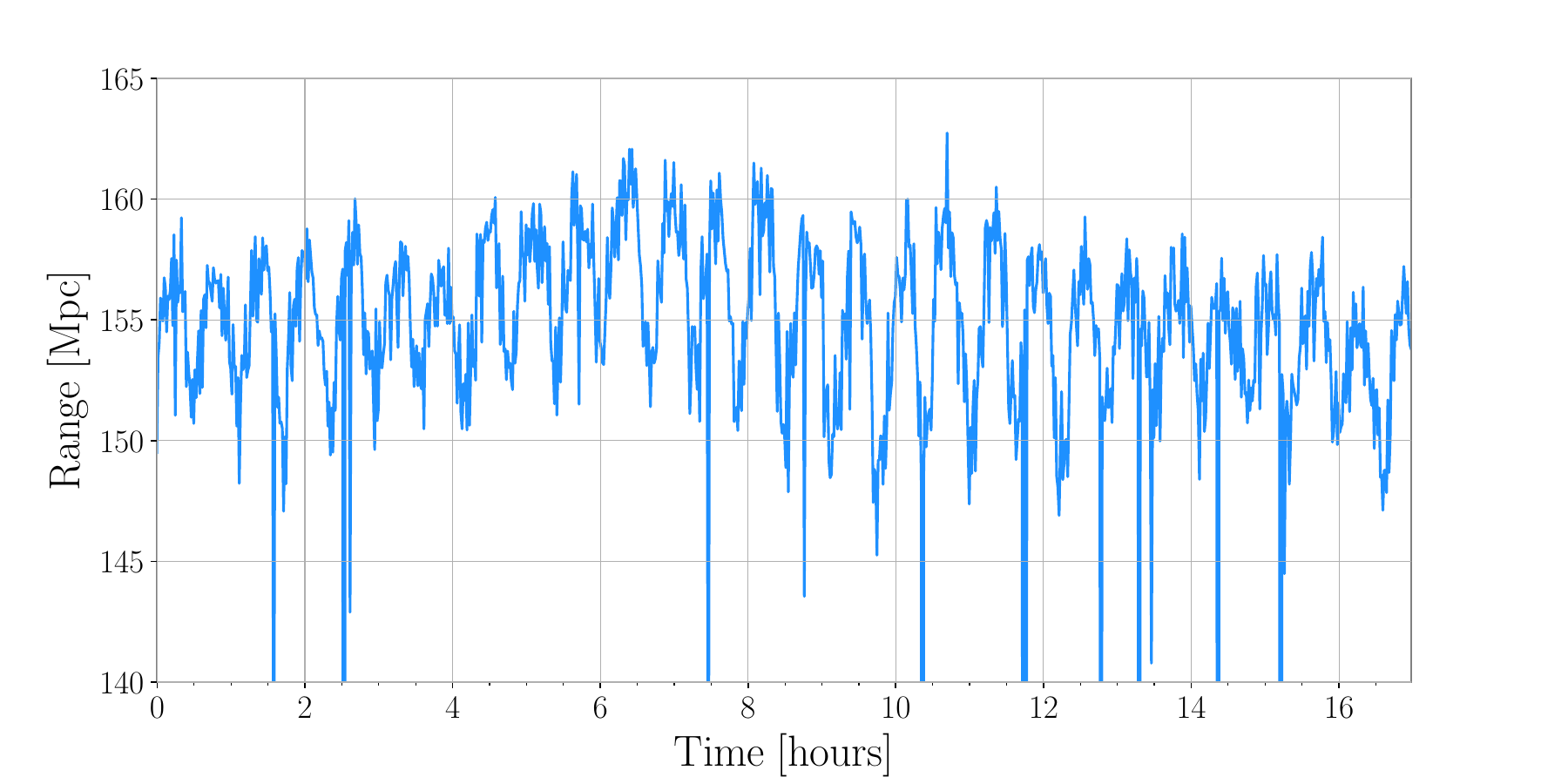}
    \caption{Range variations as observed on Dec 23 2023 at LLO. The variations are present for most of the day, and the magnitude of the drops ranges from roughly 5-10~Mpc.}
    \label{fig:range_osc}
\end{figure}

During O4a, from time to time, the observed \ac{BNS} range exhibited oscillations with a period of about 30 minutes and a range variation of about 5-15~Mpc, lasting for all or part of a day. These variations can be seen in \fref{fig:range_osc}. The range variations are the result of broadband excess noise in $h(t)$.

Searches for the cause of these oscillations identified accelerometers that seemed to witness motion that aligned with the oscillations~\cite{65146,65497,68041}. There is a line at around 30~Hz produced by the \ac{HVAC} system, and initial investigations hypothesized that this line changing amplitude could be responsible for the observed \ac{BNS} range oscillations. To check, a shaker injection at 30.5~Hz was performed on the vacuum enclosure of the chamber containing the X-arm end test mass. The 30.5~Hz shaker injection could not re-create the broadband effect we observe in $h(t)$~\cite{70525}.

The hunt for what may be causing these oscillations was continued by analyzing the output of the summary page tool Lasso. As described in~\cite{Walker:2018ylg}, Lasso can produce overlays of the \ac{BNS} range with auxiliary channels, allowing identification of channels with similar periodicity to the range variations. 
However, range variations often include secular trends combined with the oscillations, such that the channels found by Lasso often fail to align with the period of the oscillations and are thus unlikely to yield information about their cause.  

The channels that seemed to correlate the most, according to the Lasso algorithm, were primarily located at the X-arm end station and were related to temperature sensors. Additional days with similar leading channels were found -- along with channels not explicitly measuring temperature but sensitive to it~\cite{69034}. Additional investigations have mentioned issues with temperature control~\cite{69020}. While the evidence seems to implicate temperature effects at the X-arm end station, a deeper study during all of O4a could not yield the actual cause, as the oscillations are still present in O4b. The coupling mechanism between the thermal variations and $h(t)$ is not currently known.

\subsubsection{84~Hz $h(t)$ noise}
During ER15, excess noise around 84~Hz in $h(t)$ data was present. The noise would appear and disappear, suggesting that its cause might be related to dehumidifiers and fans which turn off and on. 

\begin{figure}[htp!]
    \centering
    \includegraphics[width=0.7\textwidth]{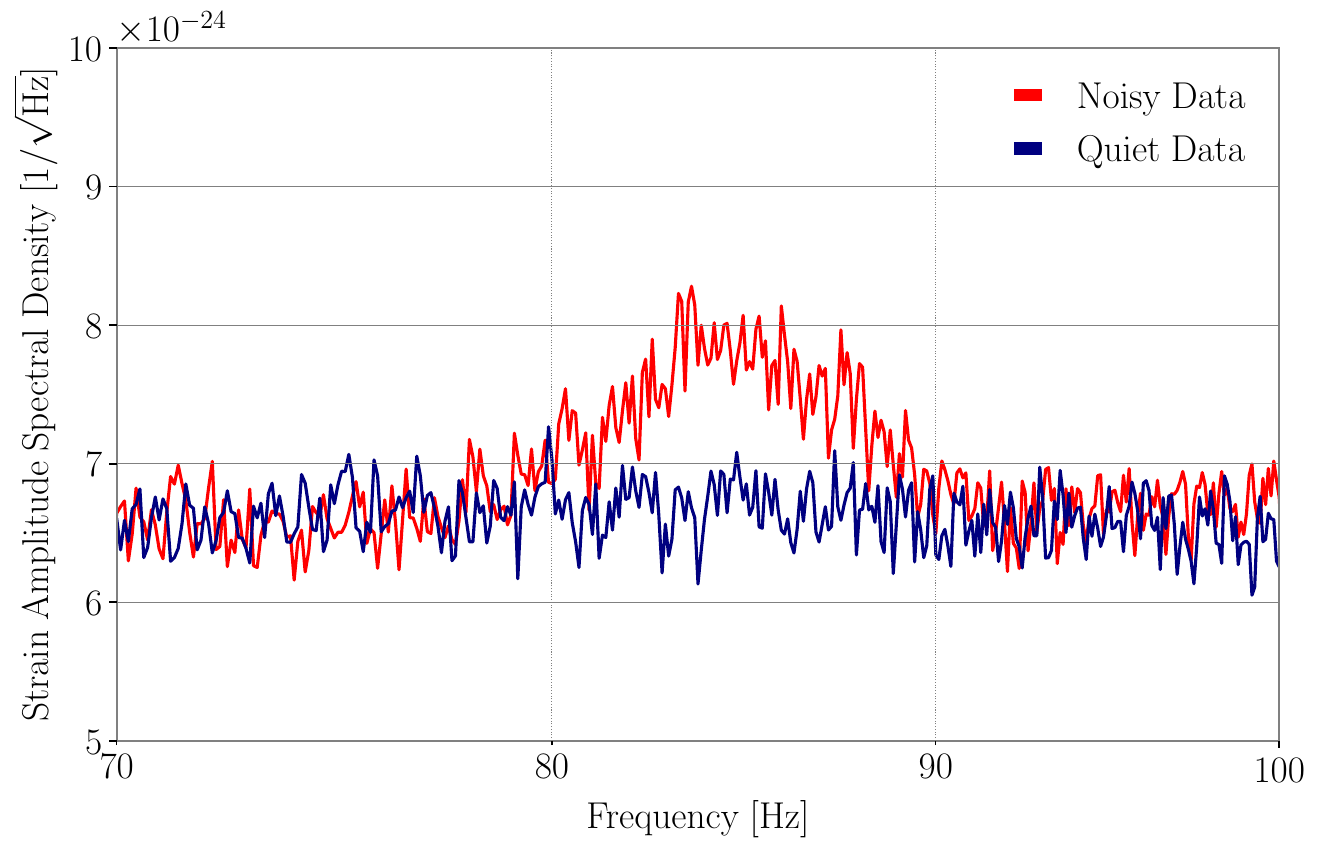}
    \caption{Comparison of data in which excess noise was present at around 84~Hz on March 2, 2023, seen here as a visible ``bump". The quiet data is taken from the same day when this noise was absent in $h(t)$.}
    \label{fig:84hzspectra}
\end{figure}

This noise was present in accelerometers at both end stations, but was louder at the Y-arm end station. Analysis of spectrograms and outputs of the channels that monitor dehumidifiers and fans pointed towards the 84~Hz source being related to two exhaust fans located at the Y-arm end station~\cite{63964}. To further confirm that this excess noise is coming from the Y-arm end station, broadband acoustic injections were done at both the X-arm end station and the Y-arm end station. These injections revealed a sharp mechanical resonance at around 84~Hz at the Y-arm end station~\cite{63399}. The exact coupling mechanism of the fans into $h(t)$ is still unknown, but the source was removed by moving the fans off the beam tube enclosure doors~\cite{64265}. 

\section{Event Validation in O4}\label{event_valid}
Validation of \ac{GW} candidates is a crucial step that enhances our confidence in the astrophysical origin of the candidate events and the reliability of source parameter-estimation results. Event Validation refers to the process of checking for the presence of any data-quality issues surrounding the time of an event and conveying this information to the relevant data analysis groups in the collaboration.
These assessments build upon the initial vetting conducted by the Rapid Response Team (RRT), a joint \ac{LVK} working group tasked with promptly responding to event candidate alerts~\cite{KAGRA:2021vkt}. 
The RRT conducts a series of prescribed data-quality checks, informed by the \ac{DQR}; see \sref{dqr}, and verifies the overall functioning of the low-latency pipeline infrastructure.
This team provides round the clock coverage, comprising rotating on-shift scientists and experts from the various areas of detector operations, including \ac{DetChar}. The prompt assessment, made after the alert for a significant event is produced by online searches, is then reconsidered more thoroughly and in more detail by the Event Validation task force to determine a final evaluation of the data-quality for every event candidate.
Multiple checks are performed on the data, e.g., making sure the detector is operating in a nominal data-taking configuration, identifying any noise artifacts that could bias source property estimation, and checking for any inconsistencies in the output of the \ac{PEM} sensors as that may suggest noise coupling between the environment and strain data.
If the data-quality around the candidate event is found to be unsatisfactory, further data processing techniques such as bayesian noise inference and transient noise removal may be applied \cite{Cornish:2020dwh}. For example, BayesWave data cleaning and linear noise subtraction were applied to a total of 17 events during O3. The catalog papers discuss these events and techniques in more detail~\cite{LIGOEVO3, LIGOScientific:2021djp, LIGOScientific:2020ibl}.

There have been a number of changes and improvements in the event validation procedure since the last observation run. These changes include:
\begin{itemize}
    \item An \ac{LVK} event validation roster for all active detectors: O4a event validation infrastructure is designed to take information from \ac{LIGO}, Virgo and KAGRA interferometers. This centralization has reduced the person power and time required to perform event validation compared to past Observing runs, during which the LIGO and Virgo Collaborations conducted validation of their detector data separately~\citep{LIGO:2021ppb,Virgo:2022ysc}. Additionally, the unified framework has ensured more standardization and uniformity in the procedures for evaluating data-quality and the tools utilized for the assessment.
    \item More automated event validation software infrastructure: The event validation infrastructure in O4 is centralized and is maintained using git version control on the event validation website, accessible to all \ac{LVK} members. 
    This centralized infrastructure allows easier information flow from DQR, to noise mitigation teams and other data analysis groups. The event validation website acts as a repository of all the details related to event validation including a list of events, the contact information of the volunteers and RRT experts, links to event’s DQR report and the event validation. For each event, an  issue page is created where any additional details regarding the event’s data-quality can be discussed. This page is also linked from the Event validation website. 
    \item More automated DQR infrastructure: The DQR is a \ac{DetChar} tool used to assess the data-quality surrounding an event time, as detailed in \sref{dqr}.
    The O4 version of the DQR has undergone significant upgrades compared to O3, incorporating automated checks that offer insights into the interferometer's state and data-quality around the event time.
    The results of these checks are displayed in the form of labels (Pass, Data Quality (DQ) issue, or Task fail), indicating whether a particular data-quality check has been successful or not.
    The event validation volunteers in O4 have made extensive use of these results for validation purposes.

\end{itemize}

\subsection{Data Quality Report}\label{dqr}

As mentioned, the primary tool used in event validation is the DQR~\cite{O4DQR}. In O3, similar DQR toolkits were separately used by the \ac{LSC}~\cite{LIGO:2021ppb} and Virgo Collaboration~\cite{Virgo:2022kwz} to evaluate candidates from the GWTC catalog. 
For O4, we used the experience gained from O3 to improve the DQR, with a focus on improving the speed and robustness of analyses, increasing the fraction of analyses that were automated, and generalizing the software to support analysis of data from all ground-based gravitational-wave detectors. 
One key upgrade was the use of a p-value, which is the probability of failing to reject the null hypothesis, to automatically identify data-quality issues that could impact the detection or analysis of gravitational-wave candidates. 
Additional details about the DQR architecture used in O4a can be found in~\cite{O4DQR}.

A wide variety of different analyses were used as part of the DQR framework during O4a.
Tests that were used to analyze \ac{LHO} and \ac{LLO} data include:
estimates of noise contributions from the observatory environment~\cite{Helmling-Cornell:2023wqe}, 
statistical correlations between strain data and auxiliary-channel information~\cite{Smith:2011an}, 
predictions of the presence of glitches using only auxiliary information~\cite{Essick:2020qpo}, 
analytic identification of excess power in spectrograms of the strain data~\cite{Vazsonyi:2022jul,Chatterji:2004qg}, 
machine-learning image classification of spectrograms of the candidate~\cite{Alvarez-Lopez:2023dmv}, 
quantitative estimates of the data stationarity~\cite{Mozzon:2020gwa}, 
estimates of the Rayleigh statistic of the data, 
measurements of the local glitch rate~\cite{doleva2022analysis}, 
and monitors of the detector range at the time of the candidate. 
These tasks were completed on two different timescales. 
Most tasks were completed within 5 minutes of a DQR being launched, allowing these tasks to be used as part of the initial rapid response to identified gravitational-wave signals. 
Additional tasks were available within a few hours to help with additional offline event validation of each candidate. 

A key feature of this updated DQR was the ability to automatically flag DQ issues in the candidate events identified by tasks based on the reported p-value. 
Candidates with DQ issues reported by the DQR underwent additional scrutiny as part of the rapid response to gravitational-wave candidates in O4a.
No additional human follow-up of the candidate data-quality was completed in low latency when no DQ issue was identified.
All candidates were further analyzed offline, however, regardless of the conclusion reached in low latency.

We found that the choice of p-value threshold strongly impacted the rate of false alarms from the DQR. 
The p-value threshold chosen to identify a data-quality issue was changed partway through the run for this reason; 
at the start of O4a, a threshold of 0.1 was used, but was eventually changed to 0.05.
This lower threshold reduced the rate of false alarms with minimal reduction in the true alarm rate. 
We also changed the set of tasks used throughout the run to improve the true alarm rate and reduce the false alarm rate. 
Using a single p-value threshold was also suboptimal, as the exact definition of the reported p-value varied between tasks.
This led to tasks either overreporting or underreporting the presence of DQ issues. 
This limitation has been addressed for O4b by introducing task-specific thresholds that are informed by our O4a experience.

\subsection{Event Validation procedure}\label{valid_procedure}

The Event Validation workflow is shown in \fref{fig:eval-flowchart}.
For each event, the volunteer (or volunteers) assigned to the week-long validation shift is (are) immediately notified. They receive all necessary information about the event, as well as instructions on how to validate it. This includes links to the Event Validation form, the DQR and the GraceDB (Gravitational-Wave Candidate Event Database)~\footnote{\href{https://gracedb.ligo.org}{www.gracedb.ligo.org}} page for the event.  In the event validation form, the validator can fill the event details for each detector. These details include the ``validation conclusion for the detector''; the options are ``Not Observing'', ``No Data Quality Issues'', and ``Data Quality Issues''. 
This information is then used as input for glitch subtraction before parameter-estimation analysis, as detailed in the next paragraphs.
Once the validator is satisfied with their findings, they can submit the validation form. This validation conclusion is then passed to the noise mitigation review team.

The noise mitigation team is responsible for assessing whether any excess power within the target time-frequency analysis window of any candidate is sufficiently non-Gaussian to require further action~\citep{Davis:2022ird,Mozzon:2020gwa}. We do this by comparing the PSD noise variance in each identified time-frequency region and check it is consistent with Gaussian noise. For regions which are not consistent with Gaussian noise (p$-$value $<$ 0.01) there are two options available. If the noise is sufficiently isolated in time and frequency the noise transient can be subtracted from the data. All noise-subtracted data in O4a were produced by the BayesWave algorithm~\citep{Cornish:2014kda,Cornish:2020dwh}. The procedure of how this is done is described in the Appendix of Ref~\citep{KAGRA:2021vkt}.


\tikzstyle{proc} = [rectangle, minimum width=2cm, minimum height=8mm, text centered, text width=3cm, draw=black]
\tikzstyle{decision} = [diamond, minimum width=2cm, minimum height=2cm, text centered, text width=15mm, draw=black]
\tikzstyle{arrow} = [thick,->,>=stealth]

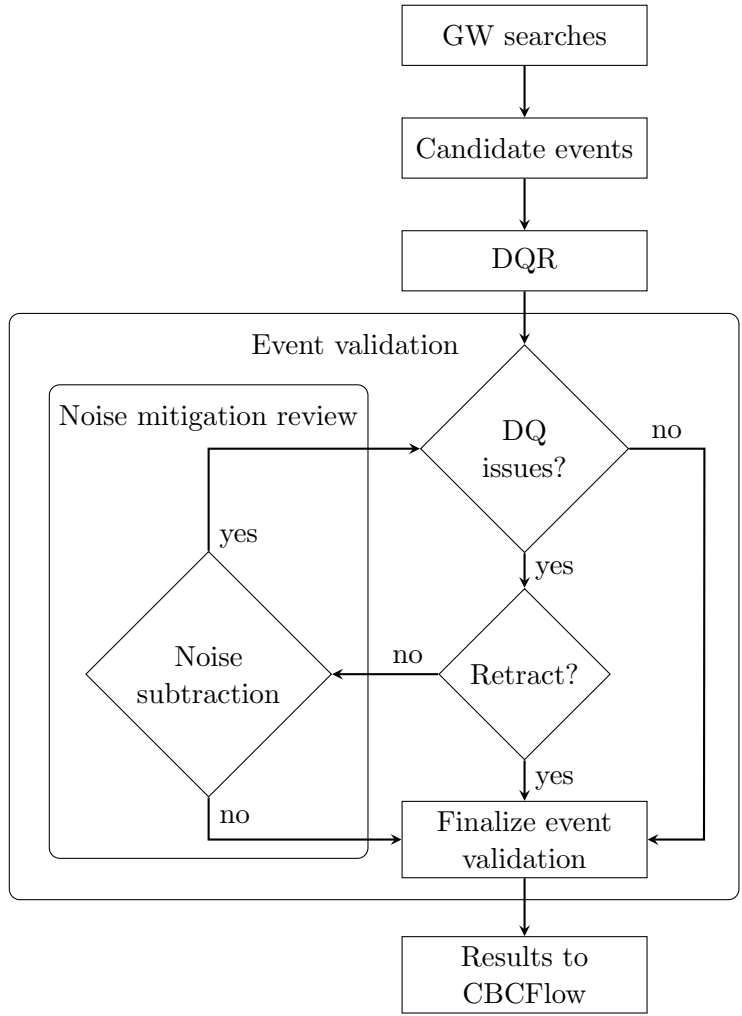
\begin{figure}[h]
    \small
    \centering
    \begin{tikzpicture}[node distance=15mm]
        \node (start) [proc] {GW searches};
        \node (events) [proc, below of=start] {Candidate events};
        \node (dqr) [proc, below of=events] {DQR};
        \node (issue) [decision, below of=dqr, yshift=-10mm] {DQ issues?};
        \node(ev)[rectangle,rounded corners, minimum width=97mm, minimum height=78mm, draw=black,anchor=north,text centered,text depth=70mm,right of=issue, xshift=-35mm, yshift=-21mm]{Event validation\hspace*{5mm}};
        \node (retract) [decision, below of=issue, yshift=-15mm] {Retract?};
        \node (sub) [decision, text width=20mm, left of=retract, xshift=-27mm] {Noise subtraction};
        \node(mitig)[rectangle,rounded corners, minimum width=38mm, minimum height=63mm, draw=black,anchor=north,text centered,text depth=55mm,above of=sub, xshift=0mm, yshift=-8mm]{Noise mitigation review};
        
        \node (fev) [proc, below of=retract, yshift=-7mm] {Finalize event validation};
    
        \node (cbc) [proc, below of=fev,yshift=-3mm] {Results to CBCFlow};
     
        \draw [arrow] (start) -- (events);
        \draw [arrow] (events) -- (dqr);
        \draw [arrow] (dqr) -- (issue);
        \draw [arrow] (fev) -- (cbc);
        \draw [arrow] (issue) -- node[anchor=west] {yes} (retract);
        \draw [arrow] (retract) -- node[anchor=west] {yes} (fev);
        \draw [arrow] (retract) -- node[anchor=south,xshift=3mm] {no} (sub);
        \draw [arrow] (sub) |- node[anchor=west,yshift=-12mm] {yes}  (issue);
        \draw [arrow] (sub) |- node[anchor=west,yshift=3mm] {no} (fev);
        \draw [arrow] (issue) -| node[anchor=south,xshift=-5mm] {no} ([shift={(10mm,0mm)}]issue.east)-- ([shift={(7.5mm,0mm)}]fev.east)--(fev); 
    \end{tikzpicture}
    \normalsize
    \caption{Event validation workflow. The DQR output is evaluated for data-quality issues. If none are found, the event is validated and the data proceeds to downstream analysis. If DQ issues are present, the event candidate may be retracted or noise subtraction recommended. The results of noise subtraction are then reassessed for residual DQ issues, repeating the evaluation process if needed.}\label{fig:eval-flowchart}
\end{figure}
     
To assess the efficacy of the noise subtracted data we compare the Gaussianity of the noise-subtracted data within the targeted time–frequency window to Gaussian
noise~\cite{Vazsonyi:2022jul}. Noise-subtracted data consistent with Gaussian
noise is deemed sufficiently stationary for parameter
estimation. If the noise is extended in time and frequency such that noise subtraction is not appropriate, or the noise subtracted data were not sufficiently stationary, the noise mitigation team can recommend restricting the time-frequency analysis window, so the parameter estimation analysis does not take into account any of the noise. The final recommended time-frequency analysis window, along with the recommended data frame name, can then be sent through CBCFlow for parameter estimation to automatically retrieve the information and start their analyses. CBCFlow is a Python library that facilitates storage and transfer of event metadata \citep{CBCFlow}.

\subsection{Validation of O4a events found by online search pipelines}\label{O4a_validation}

 In O4a, online searches generated alerts for 92 significant event candidates.
Out of these, 11 were retracted by the RRT due to evident contamination from noise artifacts or other issues that resulted in inaccurately estimated event significance, rendering their astrophysical origin improbable~\citep{2023GCN.35000....1L,2023GCN.34911....1L,2023GCN.34728....1L,2023GCN.34599....1L,2023GCN.34378....1L,2023GCN.34367....1L,2023GCN.34219....1L,2023GCN.34206....1L,2023GCN.34172....1L,2023GCN.34065....1L,2023GCN.33871....1L}.
Some of the events from the remaining 81 candidates required noise mitigation through glitch subtraction.
The procedures for assessing the necessity of glitch subtraction and its execution, along with the evaluation of the result, are detailed in~\citep{Davis:2022ird}.

For the remaining events showing data-quality issues but deemed not to require glitch subtraction, restrictions were implemented on the analyzed times and frequency bands surrounding the events. A common problem was low-frequency non-stationary noise, particularly in the lowest part of the detector sensitivity range, between 10 and 40 Hz, often caused by ground motion.
Examples of this noise can be seen in the middle and lower panels of Fig.~\ref{common-glitches}.
While GW searches and parameter estimation typically start at 20 Hz~\citep{KAGRA:2021vkt} to avoid the noise wall below 15 Hz (as shown in Fig.~\ref{O3bO4aASDs_fig}), when non-stationary noise is present, the lowest frequency may be increased to exclude the affected frequency range and avoid biases in the analysis results.

\section{Data Quality for Astrophysical Searches}\label{data_qual}

\ac{DetChar} seeks to help mitigate or eliminate identified noise sources as a top priority.  Since this is not always feasible and excess non-Gaussian noise remains in archival data even in cases where the issue was corrected, the \ac{DetChar} group prepares various data-quality products applied to astrophysical searches to reduce the impact of non-Gaussianity of the data on these searches.

\subsection{Data quality products for all searches}\label{data_qual_all_search}

As in previous observation runs~\citep{LIGO:2021ppb}, the \ac{LIGO} \ac{DetChar} group recommended that specific periods when the data is unusable due to severe data-quality issues be removed from analyzed data prior to performing astrophysical searches.  This is handled through defining segments (time periods specified by start and stop times) to be vetoed by data-quality flags.  Category 1 flags define times to be removed prior to running an analysis.  It is generally recommended that a consistent set of these flags be applied across all searches, making them relevant for all analyses described in later subsections.  Searches for gravitational waves have continued to become more sophisticated in their handling of suboptimal detector data, including utilization of noise subtraction techniques to extract gravitational-wave signals in data with noise transients present.  For this reason, \ac{LIGO} \ac{DetChar} was less aggressive about recommending periods of data for removal through defining data-quality flags in O4a than in previous runs.

Periods of non-stationarity with significant salvageable data in some frequency bands were left in place. However, there were still stretches of data when the detector was nominally supposed to be operating as an astronomical observatory during which the data were in practice un-analyzable due to severe issues sufficient to bias the PSD or the detector status otherwise being inconsistent with the detection of gravitational-waves.  The total deadtime percentage, defined as percentage of observation time removed by data-quality flags, was less than one tenth of one percent for each interferometer during O4a. Category 1 flags covered issues including:

\begin{itemize}
 \item Incorrect line subtraction for the LHO or LLO, generally at the start of lock stretches.
 \item Parametric instability mode~\cite{ParametricInstability} rung up and severely impacting data shortly before causing lockloss. This issue occurred infrequently in both \ac{LIGO} interferometers in O4a and also occurred in the third Observing run~\citep{LIGO:2021ppb}.
 \item A servo causing severe issues with squeezing \cite{LIGOO4Detector:2023wmz} at LHO.
 \item Violin modes~\cite{ViolinMode} rung up severely at LLO, in one case leading directly to distorted strain data followed by lockloss, and in another case causing issues with data calibration.
 \item Observing mode was defined incorrectly in either \ac{LIGO} interferometer early in the ER15 Engineering run prior to O4a.
 \item Observing mode was defined but $h(t)$ data was not stored permanently due to technical issues.
\end{itemize}


\subsection{Data quality for transient searches}

Searches for transient gravitational-waves cover gravitational-wave emission that will be in the detectable \ac{LIGO} frequency band for relatively short duration (sub-second to minutes).  These searches include matched-filter analyses detecting compact binary coalescences (CBCs) as well as searches for less well-modelled phenomena referred to as GW bursts.  In previous runs, both types of transient searches used Category 2 flags, which are typically shorter in duration than Category 1 and targeted the needs of specific analyses.  Due to improved confidence in gravitational-wave detection in the presence of noise, CBC searches have eliminated the use of traditional Category 2 flags, although some of these searches use other supplementary data-quality products instead as described below.  Unmodelled transient searches, which cannot rely on the characteristic chirp structure of the gravitational-wave signal for confirmation, continue to use these additional data-quality flags, but fewer kinds of flags, resulting in reduced deadtime compared to previous runs.

\subsubsection{Data quality for compact binary coalescence transient searches}

The primary data-quality products used in CBC searches were the iDQ ~\cite{Essick:2020qpo} timeseries. The iDQ pipeline produces statistical data-quality information based on auxiliary channel activity. In O4a, the Ordered Veto List (OVL) ~\cite{Essick:2013vga} algorithm was implemented within iDQ to create and rank an ensemble of vetoes for strain data triggered on auxiliary channels. The internal rank of OVL is then calibrated to probabilistic statements on the presence of a glitch by iDQ. To produce timeseries, the generated vetoes are applied to the strain data, and the probability that any time in the strain data contains a glitch is given by the highest ranked veto active at that time. These output timeseries are available for each detector in real-time and contain a number of statistics calculated by OVL for each time sample: the ranking statistic; the false-alarm probability (FAP), which is the probability that a random time with no glitch would be ranked at least as high as the current sample; the natural logarithm of the likelihood that transient noise is present ($\log L$); and a state vector for iDQ indicating the quality of the iDQ data.
 
CBC searches in O4a were performed in two different operating modes, online and offline. In online operations, CBC detection pipelines search for gravitational waves in near real-time with initial low-latency alerts sent to the public for significant detections on a timescale of seconds to minutes. In offline operations, CBC detection pipelines analyze archival data in high latency on a timescale of weeks. These offline searches have data from the entire Observing period available to them as well as additional data-quality products such as the Category 1 vetoes described previously. This makes the offline configuration of detection pipelines typically more sensitive than their low-latency counterparts, but comes at an additional computational cost and by definition cannot provide real-time alerts for astronomers.

The iDQ pipeline was also run in online and offline modes. As with CBC searches, the online configuration produced data available in near real-time to detection pipelines and data-quality experts. One detection pipeline, PyCBC Live~\cite{DalCanton:2020vpm}, integrated the iDQ FAP timeseries into their search to reject candidate gravitational-wave signals caused by glitches. This implementation discarded all gravitational-wave candidates with coalescence times within one second of any time satisfying $\mathrm{FAP}(t) < 10^{-4}$.

Offline iDQ differs from the low-latency version by having access to larger amounts of data when ranking and calibrating vetoes, allowing for more accurate estimation of their statistical properties. The log-likelihood timeseries produced by offline iDQ were used to construct data-quality flags. All of the times satisfying $ \log L(t) \geq 5$ were identified, and segments were constructed covering times from 0.25 s before to 0.25 s after each identified time. Data-quality flags were made by taking the union of all such segments. Different CBC searches may use these flags as they see fit. For example, the flags may be used as vetoes in the style of the Category 2 flags used in previous Observing runs, or they may be incorporated into a ranking statistic as in~\cite{Davis:2022cmw}.

\subsubsection{Data quality for unmodelled transient searches}

The coherent WaveBurst~(cWB) pipeline~\cite{PhysRevD.107.062002} was the primary online search algorithm for unmodelled transients, or bursts, used in O4a. 
The cWB algorithm has several modes of operation which allow it to be applied to both short duration and long duration searches for GW signals from the entire sky, and searches for signals from binary black holes, Galactic core collapse supernovae or magnetar bursts or flares.

Times of poor data-quality were removed from the burst searches through Category 1 and 2 flags as described above. All searches applied the same Category 1 flags as applied to the CBC searches, while Category 2 flags were developed by determining auxiliary data channels (those that monitor environmental or instrumental changes that are not sensitive to the effects of gravitational-waves) with high correlation to glitches that affect the burst searches.  Category 2 flags are applied primarily to the offline burst searches.  Similar to issues related to light intensity dips in O3, two Category 2 flags (one for the LHO and another for the LLO) were developed to exclude very loud glitches (usually with \ac{SNR}s $>100$)~\cite{LIGO:2021ppb}.  These flags had deadtime percentages of order a tenth of a percent for each interferometer in O4a. There were 60~Hz glitches at LLO witnessed by \ac{ESD} monitors which were used to develop an effective Category 2 flag specific to this observatory, with deadtime percentage of 0.069\% during O4a.

\subsection{Data quality for persistent searches}
Persistent gravitational-wave signals are predicted to take a variety of forms. For example, rapidly-rotating non-axisymmetric neutron stars can emit nearly monochromatic gravitational-wave signals~\cite{WETTE2023102880}, a superposition of many gravitational-wave emitters can create a broadband stationary stochastic background~\cite{galaxies10010034}, conditions in the early Universe may have generated a stochastic background signal~\cite{Caprini:2018mtu}, etc. The wide variety of possible signal models and our knowledge (or lack thereof) of waveform parameters motivates a similarly wide variety of analysis efforts~\cite{WETTE2023102880,Romano:2016dpx}. Nevertheless, these different search techniques may be served well by a relatively small number of data-quality products. Common data-quality products include information on which frequency bands are contaminated by narrowband noise lines, and cleaned strain data sets in which loud non-Gaussian transient noise has been removed. Different searches take differing approaches towards handling non-stationary noise; \ac{CW} searches tend to assign a lower weight to those time-frequency intervals of higher noise~\cite{CWReviewRiles} while stochastic searches typically remove those periods from the analysis. Below we describe the data-quality products used for persistent searches in greater detail.
\subsubsection{Data quality for continuous wave searches}

\paragraph{Self-gated $h(t)$}
Although noise transients typically do not impact \ac{CW} analyses, the cumulative effect of frequent and loud noise transients can degrade search sensitivity by effectively increasing the noise background, especially below $\sim$500~Hz~\cite{LIGO:2021ppb}. Removing these loud transient artifacts has proven useful in improving the sensitivity of broadband \ac{CW} searches. In O4, we have employed a more sophisticated algorithm to identify and remove such artifacts
, creating a new calibrated $h(t)$ dataset useful for \ac{CW} searches~\cite{O4SelfGating}.

\paragraph{Lists of narrow-band instrumental artifacts}
Most \ac{CW} searches depend on a catalog of known instrumental lines to veto spurious candidates or remove contaminated spectral bands from analysis. To produce the catalog, all lines visible in a high-resolution O4a-averaged spectrum (using Hann-windowed, 50\% overlapping \acp{FFT} of 7200-second-long data segments) are listed and evaluated. Artifacts that are confirmed to be non-astrophysical are added to a curated list and made available for searches to use, as in~\cite{O3LinesList}. 
Combs are always considered non-astrophysical because they do not align with a \ac{CW} signal model. Other lines are considered non-astrophysical when their instrumental/environmental source is known. Artifacts that do not have identified non-astrophysical causes are placed in a separate curated list, as in~\cite{O3UnvettedLinesList}. Only the confirmed non-astrophysical list is used to veto outliers, whereas the unvetted list is used for investigation purposes.


\subsubsection{Data quality for stochastic searches}
\paragraph{Non-stationarity cuts in stochastic searches.}
Stochastic searches in \ac{LVK} data are optimal under the assumption that the noise is stationary and Gaussian~\cite{galaxies10010034}. %
This is not the case in general, however, as introduced in \sref{section_intro} and described throughout \sref{ins_invs}. To mitigate these effects on stochastic analyses, we split our data into smaller segments, historically 192 s~\cite{stochasticO1,stochasticO2,stochasticO3}. The data stationarity is enforced by removing $h(t)$ segments with a standard deviation that varies by more than a chosen threshold between adjacent segments ~\cite{Renzini2024}. 

\paragraph{Auto-gating and stochastic DQ vetoes.} 
To mitigate the effect of glitches on frequency-domain non-stationarity cuts, a gating procedure is implemented to pre-process the data.  In O4a, gating in stochastic searches is handled by the \texttt{pygwb} workflow~\cite{Renzini2024} as described in~\cite{pygwb}  by multiplying the data with an inverse Planck-taper window. Periods around samples in the whitened data with an absolute value above a chosen threshold are marked for gating to remove the entirety of the glitch present in the data segment. Occasionally, the gating method implemented in \texttt{pygwb} produces extended gates ($\geq$20~s) that cover periods marked with Category 1 flags, as explained in \sref{data_qual_all_search}. This tool helps monitoring the emergence of new Category 1 flagged periods.

\begin{figure}
    \centering
    \includegraphics[width=0.9\textwidth]{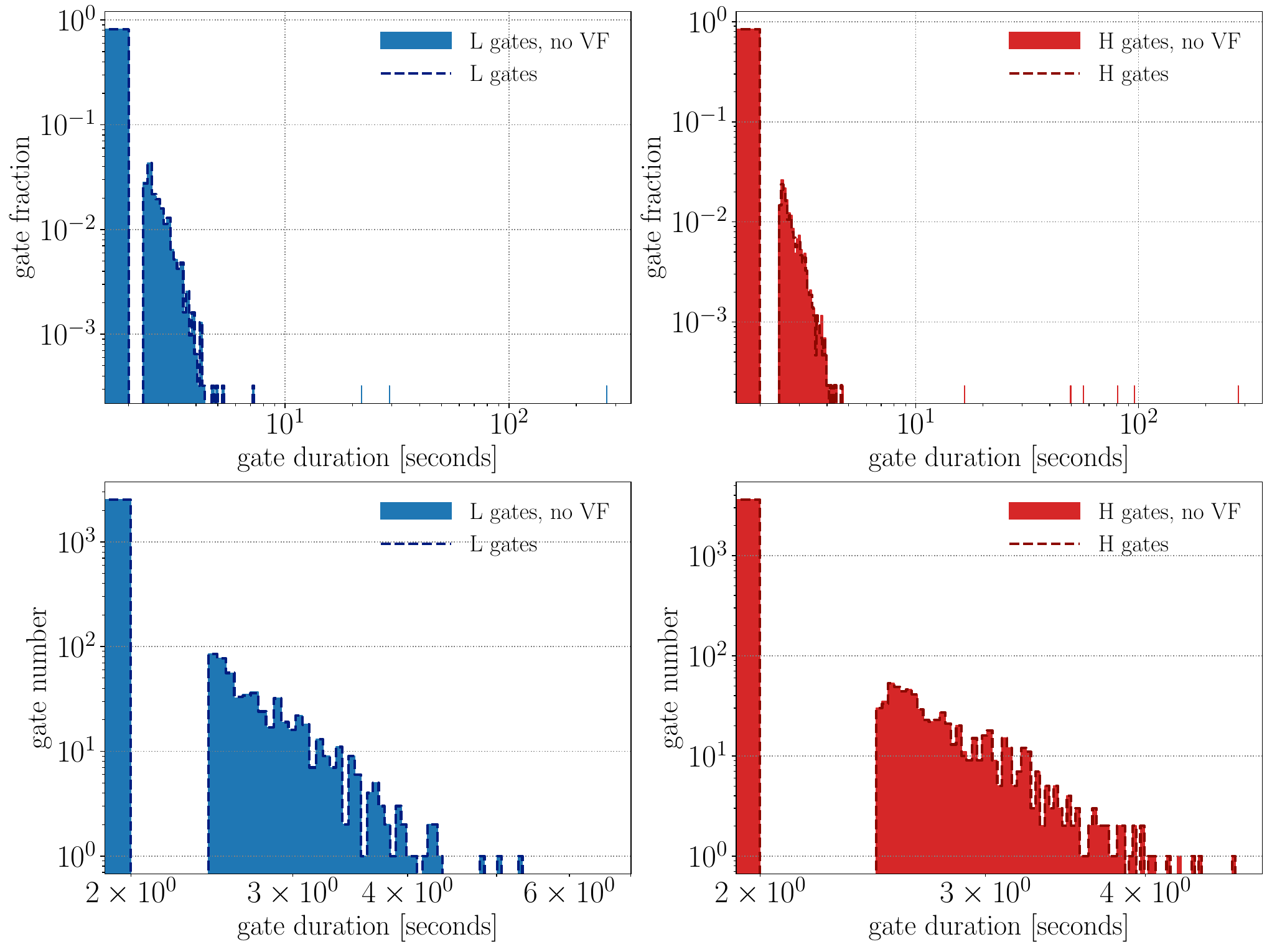}
    \caption{Effect on gate duration in stochastic data pre-processing for LLO (left-hand panels) and LHO (right-hand panels) when using the stochastic veto definer file (VF). 
    The veto definer removes a small subset of times that correspond to the (few) long-duration gates. Gates are otherwise of duration $<10$~s; $>80\%$ have duration equal to the minimum gate duration (2~s), while the rest form a cluster of gates between $[2.5, 4.5]$~s.}
    \label{fig:stochastic_gates}
\end{figure}

In O4a, there are time segments marked with Category 1 flags unique to the stochastic searches collected in a stochastic veto definer file (VF). These involve periods when a violin mode was rung up and interfering with the stationarity of the data and a specific instance when calibration lines were not being properly subtracted. The stochastic isotropic search~\cite{pygwb} was run with and without VF vetoed times to verify that the VF correctly excludes data segments that trigger long gates. In \fref{fig:stochastic_gates}, we compare gate distributions between detectors. VF effectively removed long gates in \texttt{pygwb} while maintaining overall bulk of the distribution.  Of these gates, over 80\% match the minimum gate duration (2~s); others cluster between 2.5-4.5~s.

\paragraph{Notch lists.}

As described in \sref{sec:spec-artifacts}, we monitor the coherence between $h(t)$ channels at different sites, and between $h(t)$ channel at one site and auxiliary channels at the same site. The coherence data between $h(t)$ channels indicates the frequency bins that pass the coherence threshold as shown in \fref{fig:coherence_stochmon}. They are further examined for possible instrumental causes by spectral monitor tools that also keep track of auxiliary channels (\texttt{Fscan}, \texttt{STAMP-PEM}). 
Frequency bins containing lines known to have an instrumental origin are documented and removed from the analysis.

\section{Summary and Future Prospects}\label{summ_conc}
There are multiple ways through which changes in environment impacts the \ac{GW} strain data. \ac{DetChar} group studies this interaction between the surrounding environment and the detector and develops tools to characterize and minimize the adverse impact on the \ac{GW} data quality. In this paper, we summarize the work of the \ac{DetChar} group between the end of the third Observing run and the end of the first half of the fourth Observing run. These efforts led to a factor of $\sim$50 reduction in the rate of daytime laser light scattering at \ac{LLO} as detailed in \ref{acbres}, identification of the high-frequency magnetic noise coupling at \ac{LHO}, identification of several persistent narrow-band noise features in \ac{LLO} and \ac{LHO}, a more comprehensive Event Validation and \ac{DQR} framework and an overall better understanding of the noise characteristics for astrophysical searches. 

As LIGO detectors become more sensitive, detector characterization will become more challenging. Increased sensitivity translates to higher rate of events, but could also lead to an increase in glitch rate. Our work entails not just glitch characterization and reduction, but also validation of the data quality surrounding an event. %
Lower detector noise across the band also implies higher sensitivity to persistent signals as well as narrow-band, broad-band, and/or correlated terrestrial noise sources. %
Continuous monitoring of the data quality, identification of potential noise couplings in the detector, and improvement in our software tools are some of the prerequisites for timely dissemination of robust, and accurate astrophysical results. This would, among other things, require more person power, increased automation of tools such as \ac{DQR} and Event Validation, and stronger collaboration between instrument science and \ac{DetChar} group. These efforts will lead to more robust identification of weak astrophysical GW signals in noisy LIGO data, and thus to a deeper probe of the GW sky.

Data-quality products described in this paper from previous Observing runs have been publicly released on the Gravitational Wave Open Science Center (GWOSC) webpage \footnote{\href{https://gwosc.org}{www.gwosc.org}}, and when the O4a data are publicly released, data-quality products will be released alongside~\cite{GWOSC-2023}.

\ack{
We would like to thank Christopher Berry, Jess McIver and Alan Weinstein for their many helpful comments and suggestions.
This material is based upon work supported by NSF's LIGO Laboratory which is a major facility fully funded by the National Science Foundation.
LIGO was constructed by the California Institute of Technology and Massachusetts Institute of Technology with funding from the National Science Foundation and operates under Cooperative Agreement PHY-1764464. Advanced LIGO was built under grant No. PHY-0823459. This work uses the LIGO computing clusters and data from the Advanced LIGO detectors. This document has been assigned LIGO-number LIGO-P2400320.}

\vspace{5em}
\newpage
\bibliographystyle{my-iopart-num}
\bibliography{sample.bib}

\end{document}